\documentclass[twocolumn,apj]{aastex62}

\def\cgs{{ erg cm$^{-2}$ s$^{-1}$}}

\def\apj{ApJ}
\def\aap{A\&A}

\def\apjl{ApJL}
\def\apjs{ApJS}

\shortauthors{Rajagopal et al.}

\usepackage{natbib}
\usepackage{float}
\usepackage{color}
\usepackage{graphicx}
\usepackage{textcomp,gensymb}
\usepackage{array}
\usepackage{enumitem}
\usepackage{longtable}
\usepackage{hyperref}
\usepackage{ulem}
\usepackage[bottom]{footmisc}
\usepackage{balance}

\begin{document}
\title{Identifying the 3FHL catalog: V. Results of the CTIO-COSMOS optical spectroscopy campaign 2019}
\email{changar@clemson.edu}
\author[0000-0002-8979-5254]{M. Rajagopal}
\affiliation{Department of Physics and Astronomy, Clemson University, Clemson, SC 29634, USA}
\author{S. Marchesi}
\affiliation{INAF $-$ Osservatorio di Astrofisica e Scienza dello Spazio di Bologna, Via Piero Gobetti, 93/3, 40129, Bologna, Italy}
\affiliation{Department of Physics and Astronomy, Clemson University, Clemson, SC 29634, USA}
\author{ A. Kaur} 
\affil{Department of Astronomy and Astrophysics, 525 Davey Lab, Pennsylvania State University, University Park, 16802, USA}
\author{ A. Dom\'{i}nguez}
\affil{IPARCOS and Department of EMFTEL, Universidad Complutense de Madrid, E-28040 Madrid, Spain}
\author{R. Silver}
\affiliation{Department of Physics and Astronomy, Clemson University, Clemson, SC 29634, USA}
\author{M. Ajello} 
\affiliation{Department of Physics and Astronomy, Clemson University, Clemson, SC 29634, USA}

\begin{abstract}
As a follow-up of the optical spectroscopic campaign aimed at achieving completeness in the Third Catalog of Hard Fermi-LAT Sources (3FHL), we present here the results of a sample of 28 blazars of uncertain type observed using the 4\,m telescope at Cerro Tololo Inter-American Observatory (CTIO) in Chile. Out of these 28 sources, we find that 25 are BL Lacertae objects (BL Lacs) and 3 are Flat Spectrum Radio Quasars (FSRQs). We measure redshifts or lower limits for 16 of these blazars, whereas it is observed that the 12 remaining blazars have featureless optical spectra. These results are part of a more extended campaign of optical spectroscopy follow up of 3FHL blazars, where until now 51 blazars of uncertain type have been classified into BL Lac and FSRQ categories. Further, this campaign has resulted in redshift measurements and lower limits for 15 of these sources. Our results contribute towards attaining a complete sample of blazars above 10 GeV, which then will be crucial in extending our knowledge on blazar emission mechanisms and the extragalactic background light.
\end{abstract}


\accepted{: \apj}

\section{Introduction}
The Large Area Telescope (LAT; \citealp{atwood09}) on board the {\it Fermi Gamma-ray Space Telescope} has enabled great strides in the understanding of blazars, the most powerful class of active galactic nuclei (AGN). Blazars make up the majority of the astrophysical gamma-ray source population \citep{4fgl}. The broadband spectral energy distribution (SED) of blazars is generally characterized by two distinct bumps, which are attributed to synchrotron radiation at lower energies (infrared to X-rays) and inverse Compton scattering at high energies (X-rays to $\gamma$-rays, \citealp{maraschi1994,abdo2011}). 

The Third Catalog of Hard Fermi-LAT Sources (3FHL; \citealp{ajello17}) reported 1556 sources detected in the 10 GeV -- 2 TeV energy range during the first seven years of \textit{Fermi}-LAT operation. Blazars represent the majority ($\sim$1300/1556) of these 3FHL sources. However, $\sim$45\% (578) of these blazars lack a redshift measurement and $\sim$21\% (232) are of uncertain type, leaving the 3FHL catalog highly incomplete.

Accurate redshift determinations are essential for studying numerous topics, including blazar emission processes and energetics (e.g., \citealp{ghisellini17,vandenberg}), the role of blazars in cosmic ray acceleration (e.g., \citealp{furniss13}), the properties and evolution of extragalactic background light (EBL, i.e., the integrated emission from all stars and galaxies in the universe since the reionization epoch; e.g., \citealp{abramowski12,ackermann12,dom2015,fermi2018,abhi2019}), the derivation of cosmological parameters using gamma rays (e.g., \citealp{dom2019}), and studying and understanding the cosmic evolution of blazars (e.g., \citealp{ajello14,paliya2020}).

To this end, spectroscopic campaigns of blazars have been effectively conducted with 4\,m, 8\,m and 10\,m class telescopes to investigate the nature of these sources as well as to provide accurate redshift measurements (e.g., \citealp{sbaruffati05, shaw12, massaro14, paggi14, landoni15, ricci15, alvarez16a, alvarez16b, paiano17, pena17, marchesi18, desai2019, Paiano_2019, paiano2020,pena2020}). The successes of these campaigns prove that the listed classes of telescopes are capable of distinguishing the two blazar subclasses, namely, BL Lacertae objects (BL Lacs) and Flat Spectrum Radio Quasars (FSRQs). 

FSRQs show broad emission features in their optical spectra and have higher average luminosities than BL Lacs, thus allowing easy measurement of their redshifts, while BL Lac spectra exhibit weak or very narrow emission lines (equivalent width, EW $\leq$ 5\,\AA; \citealp{urry95,marcha96}). 

Blazars are further subdivided into three categories based on the position of their synchrotron peak frequency, $\nu_{pk}^{sy}$. Sources with $\nu_{pk}^{sy} < 10^{14}$\, Hz are classified as low-synchrotron peak blazars (LSP), sources with $10^{14} < \nu_{pk}^{sy} < 10^{15}$\, Hz are known as intermediate-synchrotron peak blazars (ISP), and sources with $\nu_{pk}^{sy} > 10^{15}$\, Hz are categorized as high-synchrotron peak blazars (HSP). In the {\it Fermi} catalogs, BCU is a term used for classification for associated counterparts that (a) show blazar-like multi-frequency behavior or (b) are classified as blazars of uncertain type in the Roma-BZCAT \citep{bzcat}.
The 3FHL contains 232 BCUs, 28 of which were targeted in this work. Spectroscopic observations of such sources will enable us to classify these sources and potentially measure their redshifts.\\  

In this campaign, so far 51 BCUs have been classified into BL Lac and FSRQ categories using optical spectroscopy. Spectroscopic redshifts were also obtained for 15 of these sources (\citealp{marchesi18,desai2019}). In addition, machine learning algorithms have provided tentative classification for an additional 51 sources (\citealp{Kaur18,Silver2020}).

In this work, we report the results from spectroscopic observations of 3FHL BCUs performed using the Cerro Tololo Ohio State Multi-Object Spectrograph (COSMOS) mounted on the 4\,m Blanco telescope located at the CTIO facility. The organization of this paper is as follows: in Section~\ref{sample}, we report the criteria used for the sample selection. In Section~\ref{analysis}, we describe the observation methodology and the data analysis approach. In Section~\ref{results}, we summarize the results of this campaign, with additional consideration given to sources that exhibit spectral features, while in Section~\ref{conc}, we present our conclusions.

\section{Sample Selection \label{sample}}

In this work, we report observations of 28 sources performed with CTIO in 2019, constituting the third part of this campaign\footnote{These observations were awarded as part of the Fermi-GI cycle 10, accepted proposal 101287, PI: S. Marchesi.}. These sources were selected from BCUs in the 3FHL catalog based on the following criteria:
\begin{enumerate}
\item Lack of a redshift.
\item Optical magnitude in the V band  V$\leq$21.5.
\item Being bright in 10 GeV -- 2 TeV band: ($f_{3FHL}> 10^{-12}$ erg s$^{-1}$ cm$^{-2}$). This condition ensures that the completeness of the 3FHL catalog moves to lower fluxes as more optical observations are performed.
\item Being observable from Cerro Tololo ($-$80$^\circ<\, \text{decl.} < 20^\circ$) , above an altitude of 40$^\circ$ and during our scheduled observations from April to June.
\end{enumerate}

Details of all sources analyzed here have been summarized in Table~\ref{tab:tabulated1}. Optical magnitudes have been obtained from the 3FHL catalog (where the authors used USNO\footnote{\url{https://www.usno.navy.mil/USNO/astrometry/optical-IR-prod}} or SDSS\footnote{\url{http://www.sdss3.org/}} archival data).

\section{Observations and Data Reduction} \label{analysis}
All the source spectra for this work were obtained using the Cerro Tololo Ohio State Multi-Object Spectrograph (COSMOS) mounted on the 4\,m Blanco telescope located at the CTIO facility in Chile. The instrument was configured with the red grism and 0.9\arcsec slit, to cover a wavelength range of $\lambda =$ [5000$-$9000] \AA, corresponding to a dispersion of $\sim$4 \AA \, pixel$^{-1}$ and a spectral resolution of R $\sim$ 2100. Wavelength calibration was performed with a precision of 0.04 \AA. The fluctuations were found to be both positive and negative indicating a non-biased wavelength calibration process. The observations were performed over 3 nights in April 2019 and over 4 nights in June 2019. However, due to poor weather conditions, data from 6 total nights was used in our analysis. The observing dates for all sources along with the exposure times and V-band magnitudes have been reported in Table~\ref{tab:tabulated1}. 

Spectra for each source were obtained in triplicate or more, with varying exposure times, and the individual observations were combined to reduce the effects of instrumental noise and cosmic-ray contamination. Data reduction was performed using the standard IRAF pipeline \citep{tody86} including bias subtraction and correction for bad pixels. Flat-field normalization was also performed in order to remove any wavelength-dependent variations present in the flat-field source (but not in the observed spectrum). The reliability of the emission and absorption lines was obtained by comparing the features in each individual exposure. Additionally, all spectra were visually inspected to remove any artificial features. 

Wavelength calibration for each source was performed using the Mercury$-$Neon (Hg-Ne) lamp. A lamp spectrum was obtained after each observation of a source to account for the possible shifts in the pixel$-\lambda$ calibration, due to changes in the telescope position throughout the night. The spectra were independently inspected for emission and absorption features typical of those observed in BL Lac and FSRQ spectra (e.g., see \citealp{stickel1991,marcha1996opt,bade98,Shaw_2013}). The presence of only absorption features in a spectrum indicates either absorption by the intervening medium (along the line of sight) or by the host galaxy (e.g., Ca {\sc ii}, G$-$band, NaD, Mg {\sc i}, Ca$+$Fe lines). This enables us to set a spectroscopic lower limit of the redshift. All spectra were flux calibrated using a spectrophotometric standard. The standards were observed twice each night: one at the beginning of the night and one at the end, under the same observing conditions as the rest of the analysis. Finally, each source spectrum was corrected for Galactic reddening, using the $E(B-V)$ value obtained from NASA/IPAC Infrared Science Archive online tool\footnote{\url{https://irsa.ipac.caltech.edu/applications/DUST/}}, based on \cite{schlafly11}.
\begin{figure*}
\caption{Normalized optical spectra of the observed candidates after performing flux calibration and dereddening. The atmospheric features are denoted by $\earth$ while the absorption or emission features are labeled as per the lines they signify. \textbf{Flux is in the units of erg cm$^{-2}$ s$^{-1}$ \AA$^{-1}$.}}
\vspace{-1.5cm}
\begin{minipage}[b]{0.5\textwidth}
  \centering
  \includegraphics[width=0.97\textwidth]{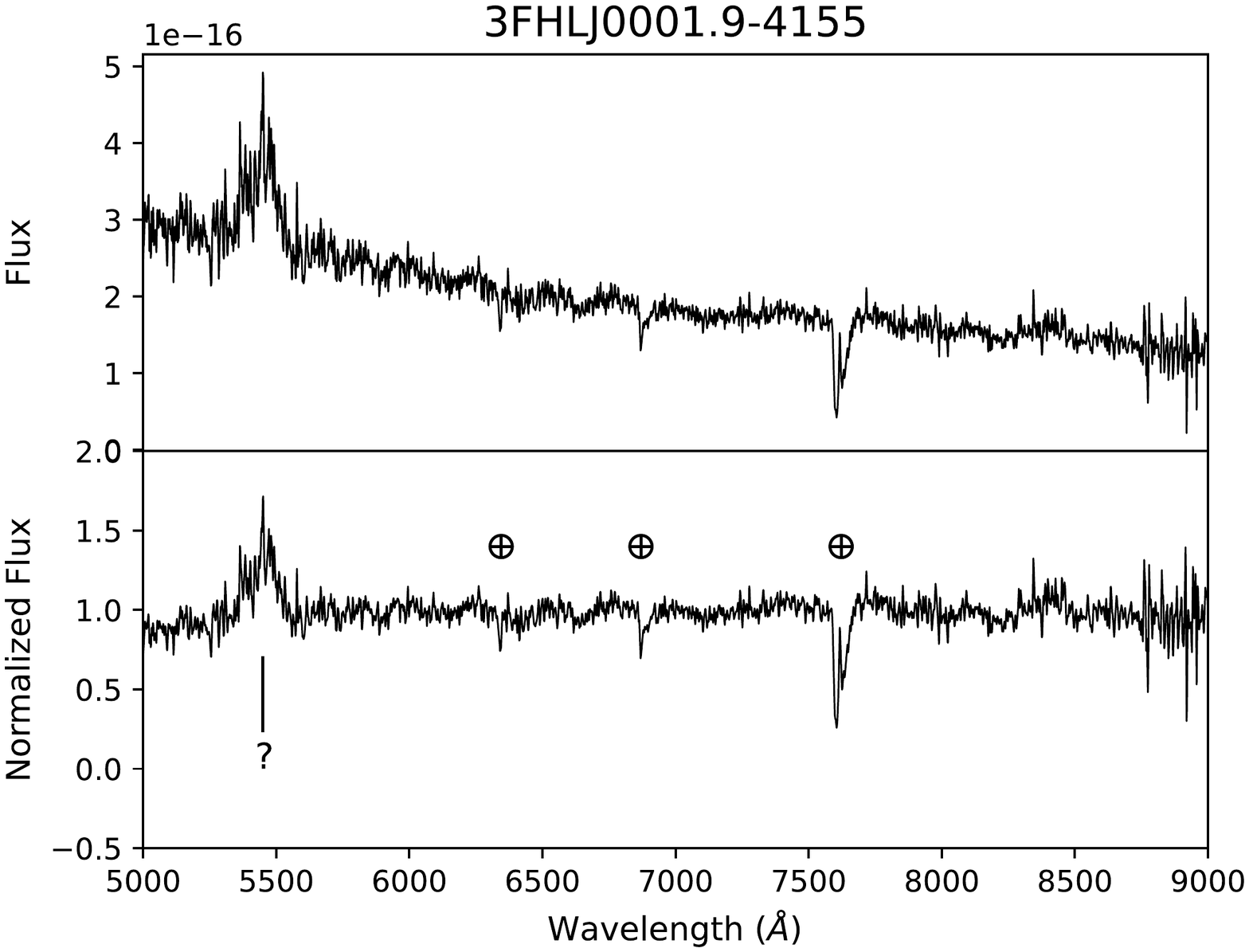}
  \end{minipage}
\begin{minipage}[b]{.5\textwidth}
  \centering
  \includegraphics[width=0.97\textwidth]{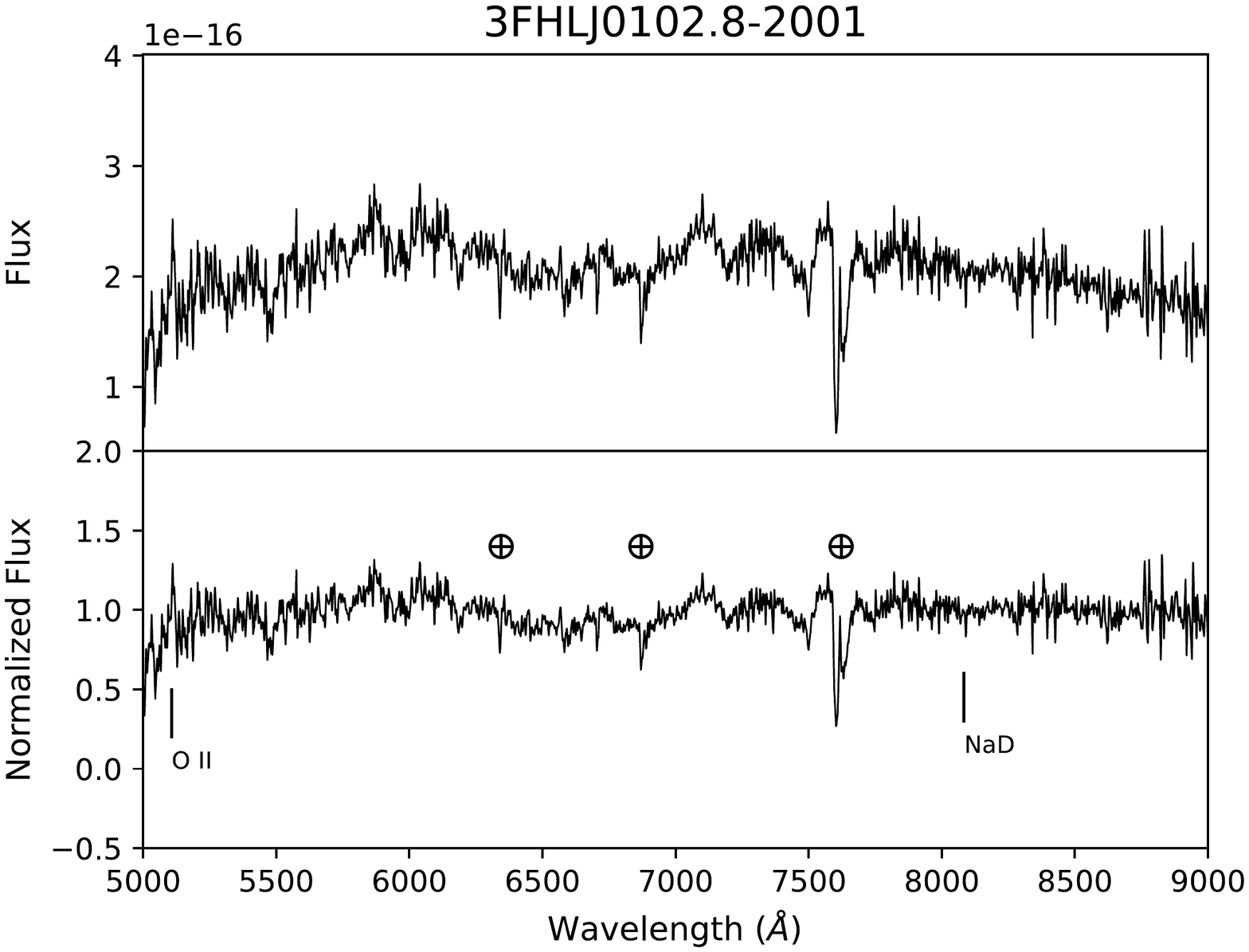}
  \end{minipage}
\begin{minipage}[b]{.5\textwidth}
\vspace{-3.7cm}
  \centering
  \includegraphics[width=0.97\textwidth]{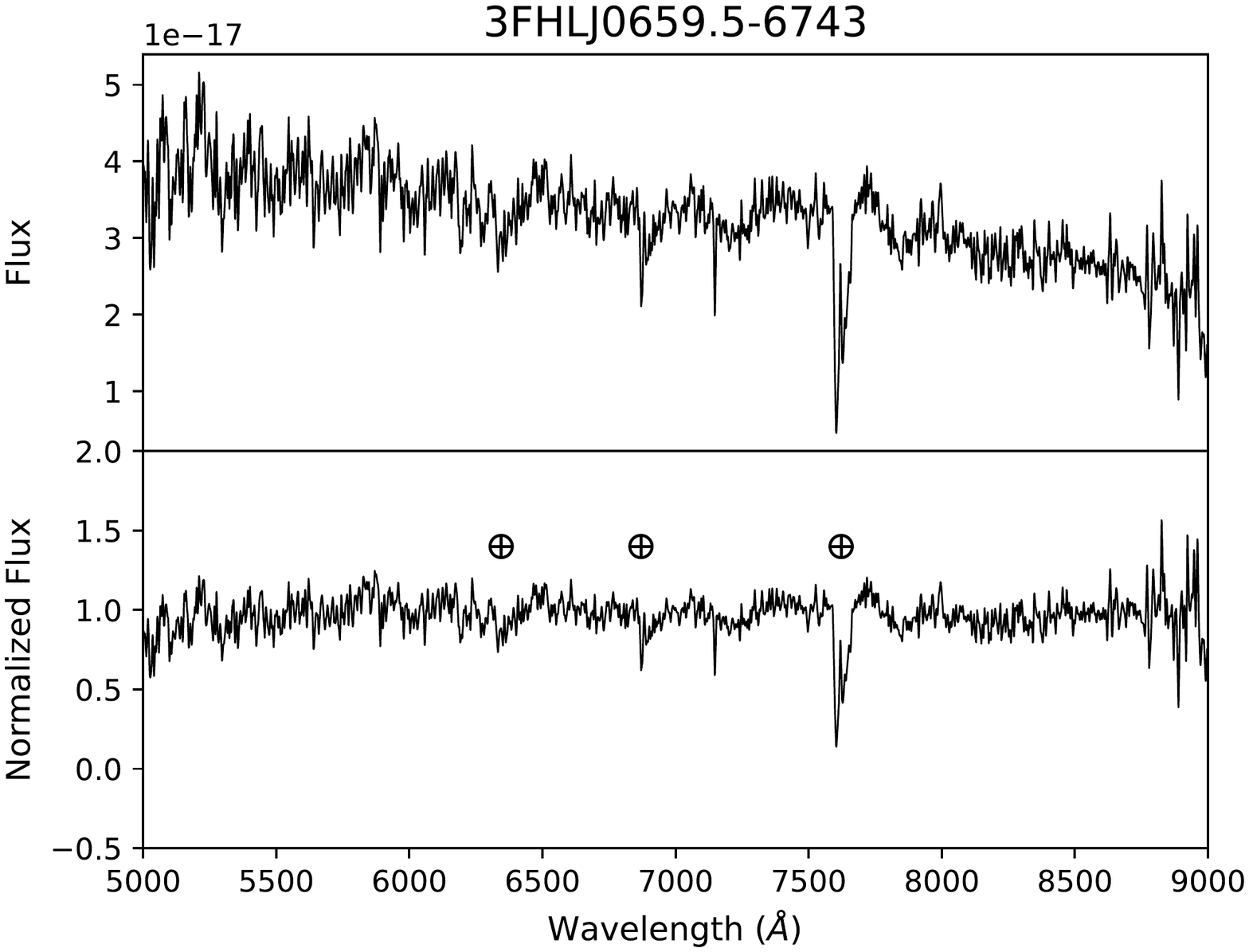}
  \end{minipage}
\begin{minipage}[b]{.5\textwidth}
\vspace{-3.7cm}
  \centering
  \includegraphics[width=0.97\textwidth]{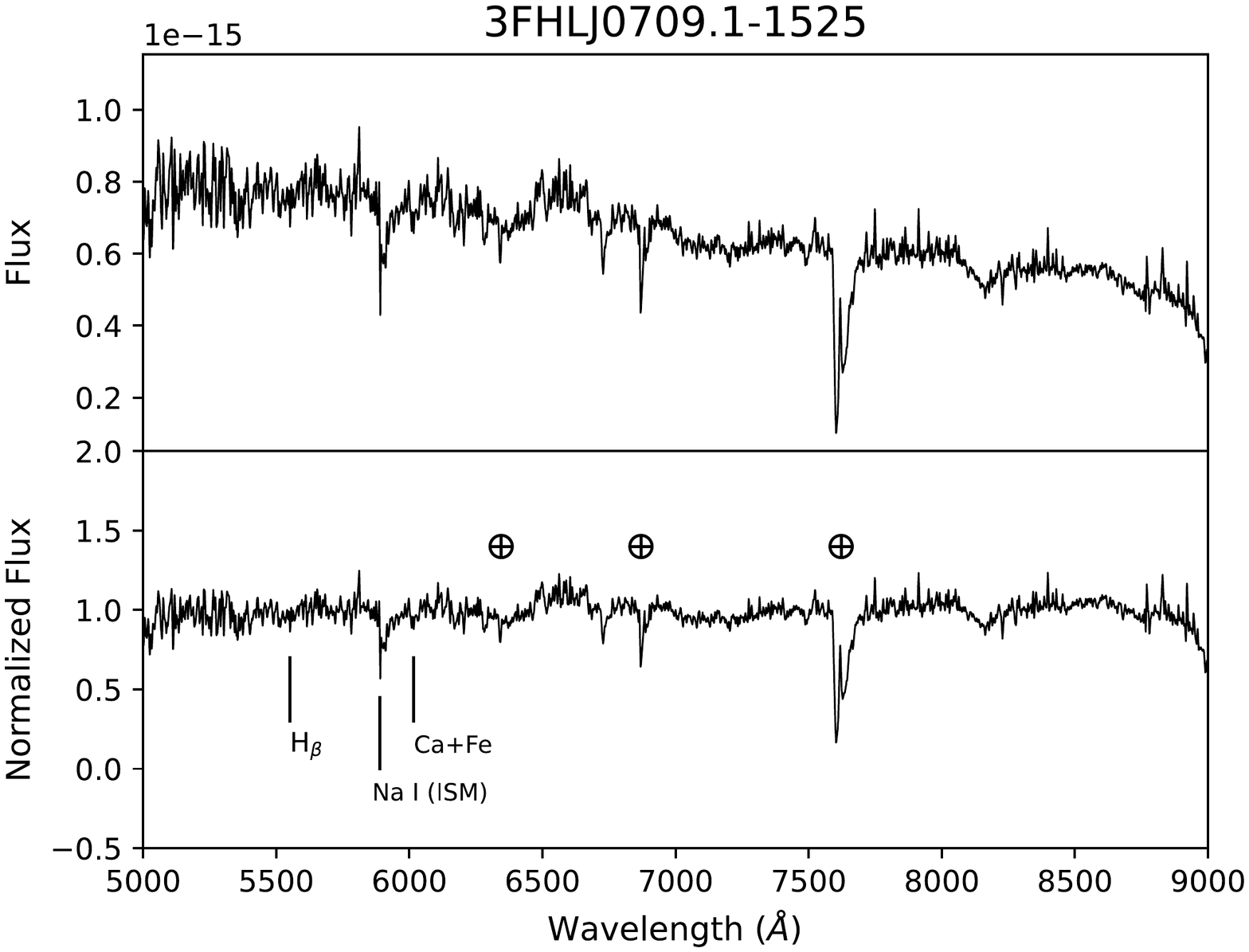}
  \end{minipage}
\begin{minipage}[b]{.5\textwidth}
\vspace{-3.7cm}
  \centering
  \includegraphics[width=0.97\textwidth]{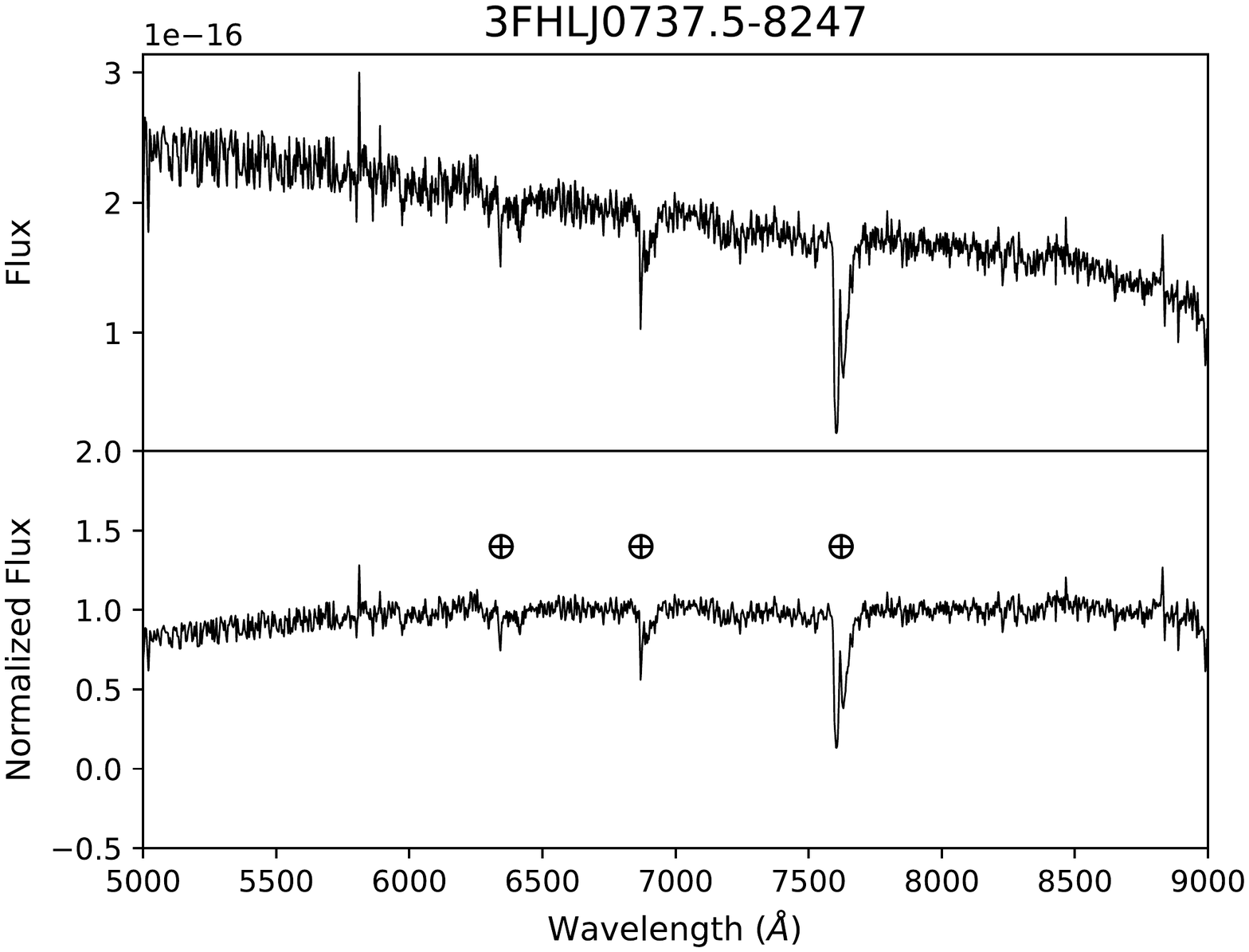}
  \end{minipage}
\begin{minipage}[b]{.5\textwidth}
\vspace{-3.7cm}
  \centering
  \includegraphics[width=0.97\textwidth]{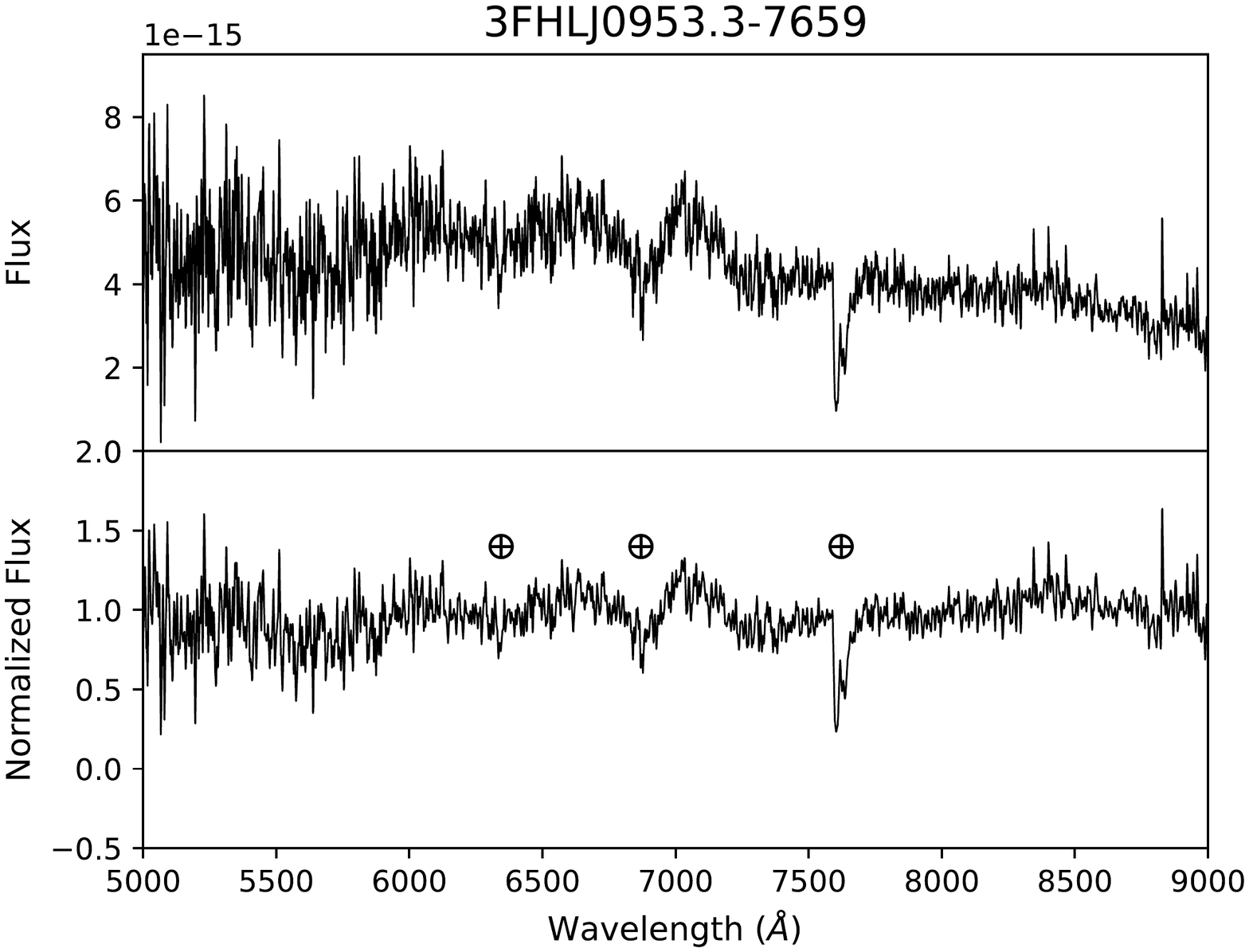}
  \end{minipage}
\end{figure*}

\begin{figure*}
\vspace{-1.7cm}
  \begin{minipage}[b]{.5\textwidth}
  \centering
  \includegraphics[width=0.97\textwidth]{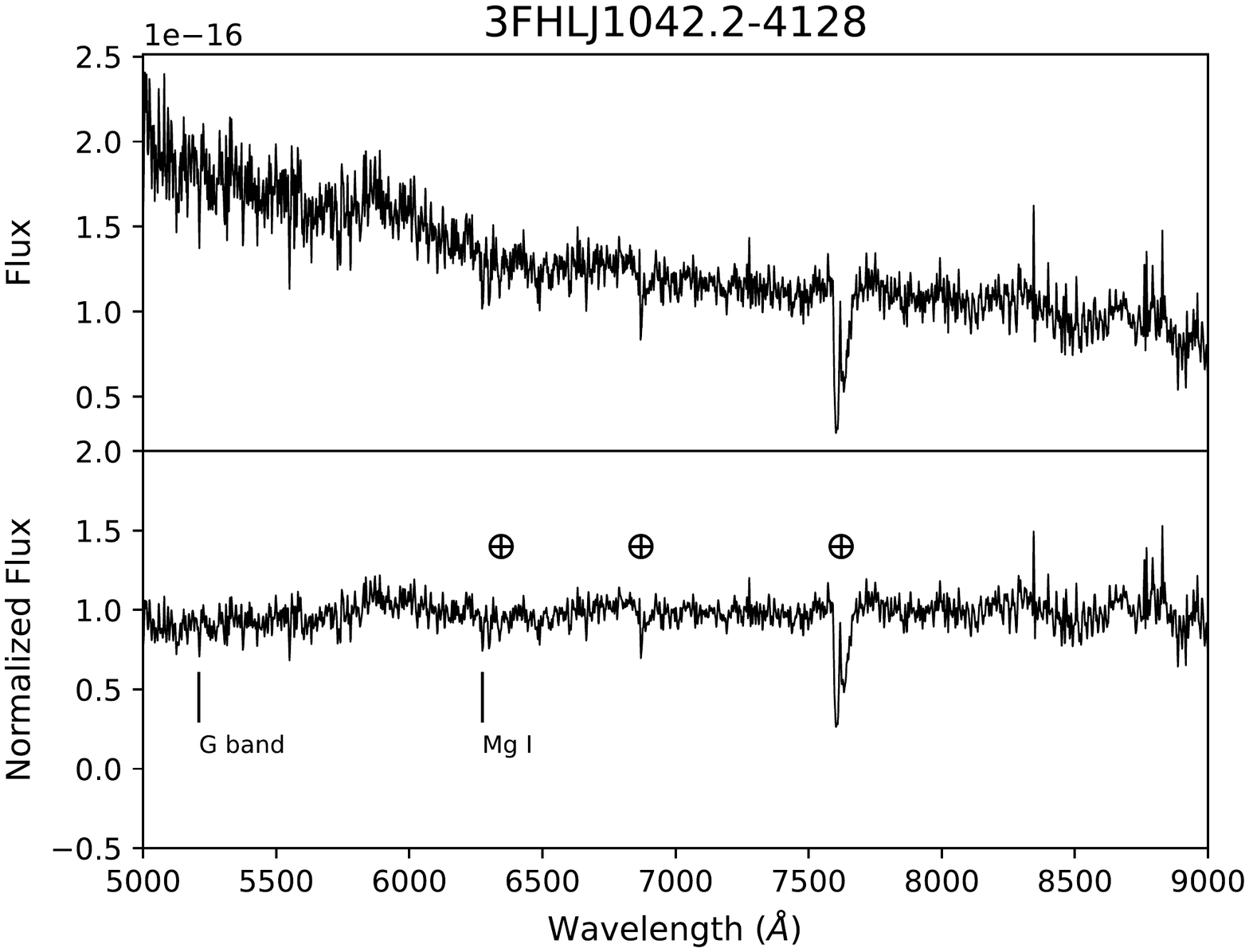}
  \end{minipage}
  \begin{minipage}[b]{.5\textwidth}
  \centering
  \includegraphics[width=0.97\textwidth]{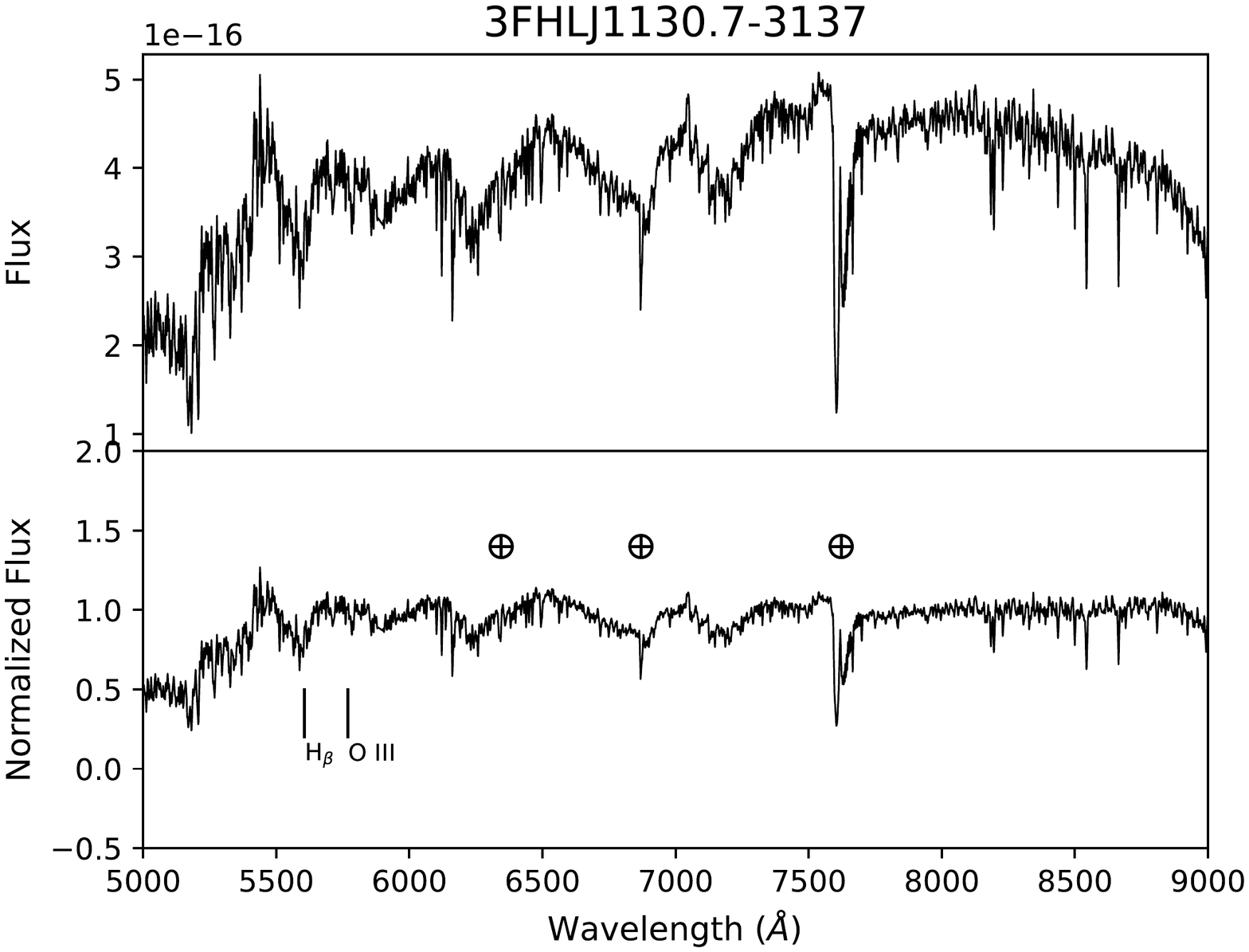}
  \end{minipage}
\label{fig:spec1}
\begin{minipage}[b]{.5\textwidth}
\vspace{-3.7cm}
  \centering
  \includegraphics[width=0.97\textwidth]{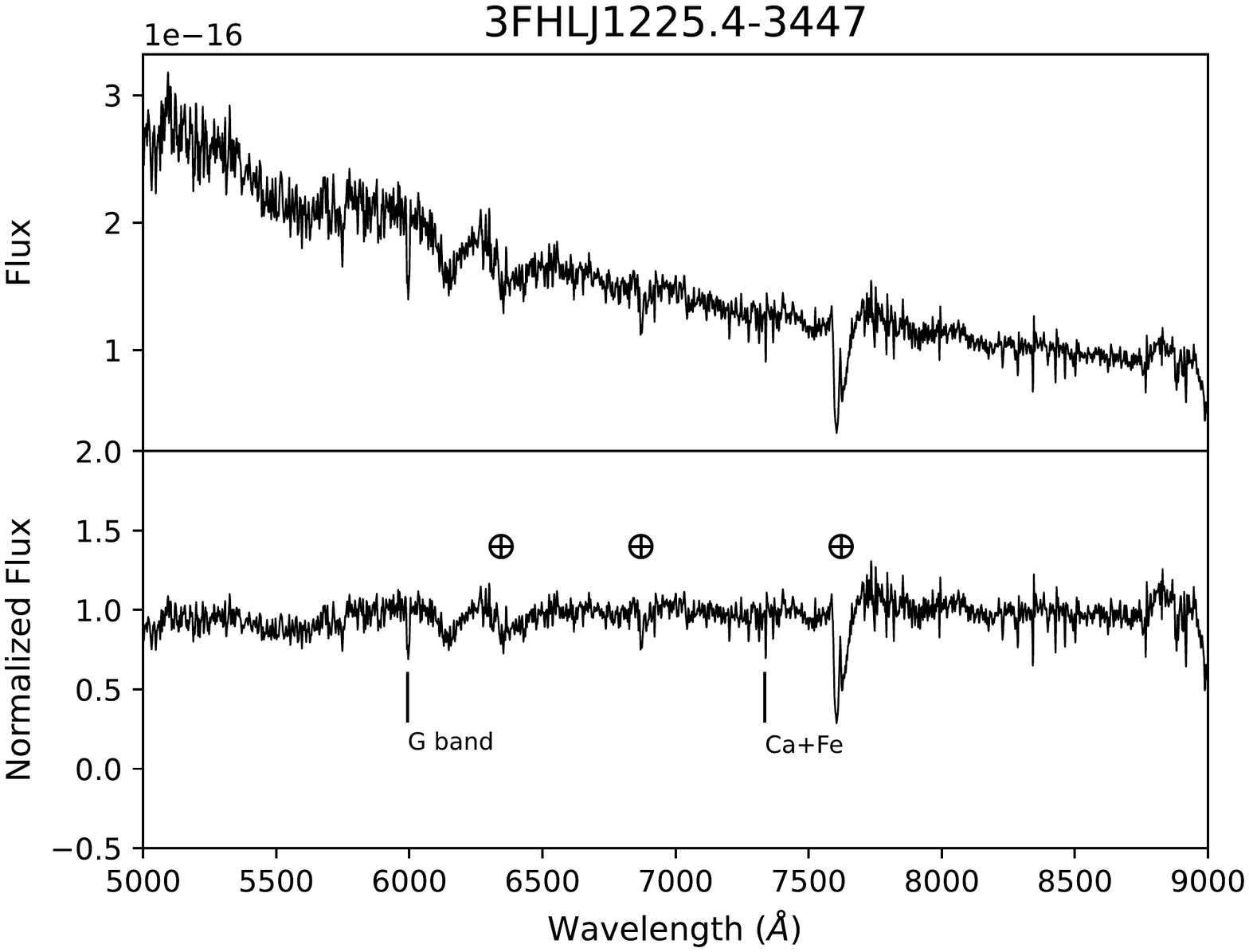}
  \end{minipage}
\begin{minipage}[b]{.5\textwidth}
\vspace{-3.7cm}
  \centering
 \includegraphics[width=0.97\textwidth]{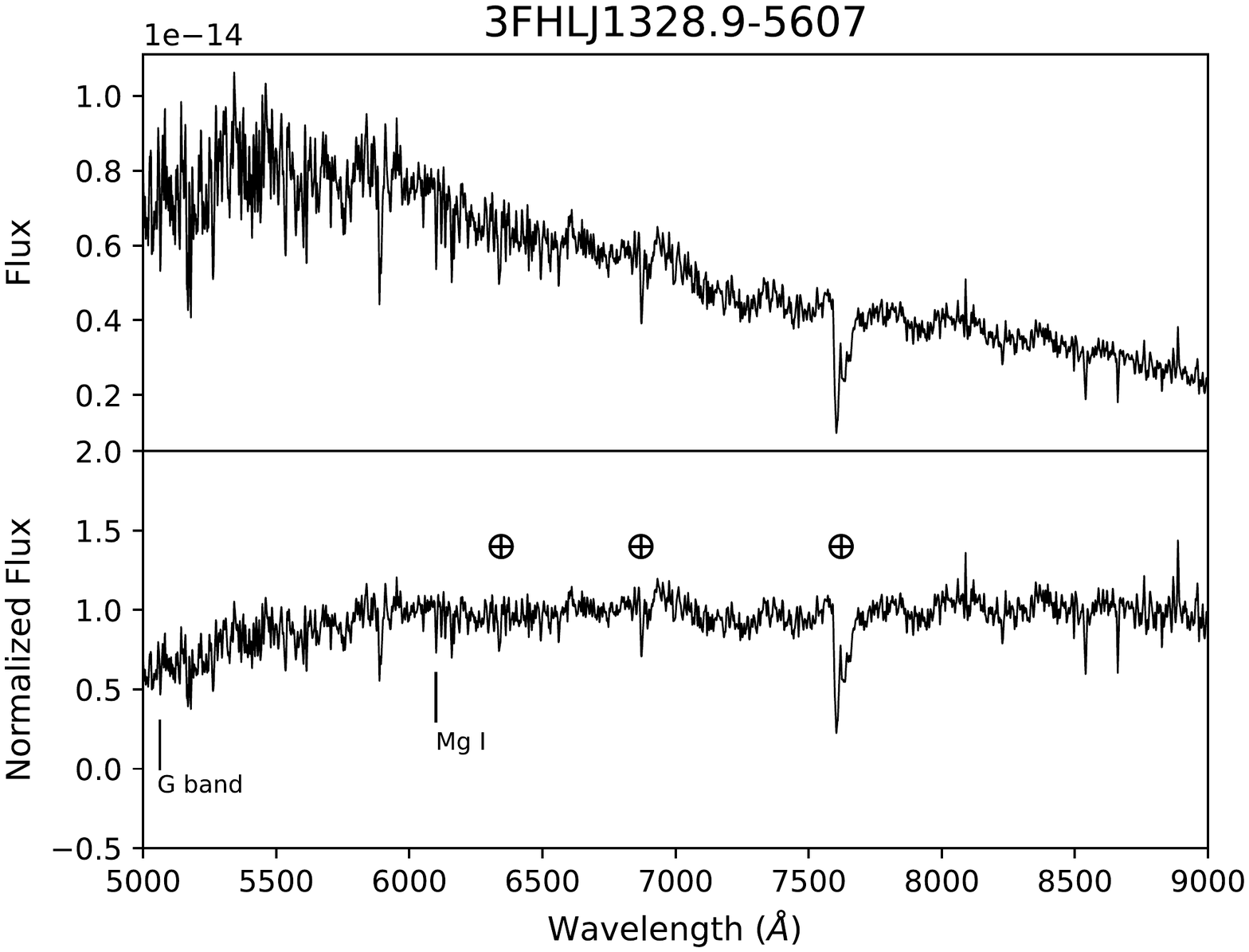}
  \end{minipage}
\begin{minipage}[b]{.5\textwidth}
\vspace{-3.7cm}
  \centering
 \includegraphics[width=0.97\textwidth]{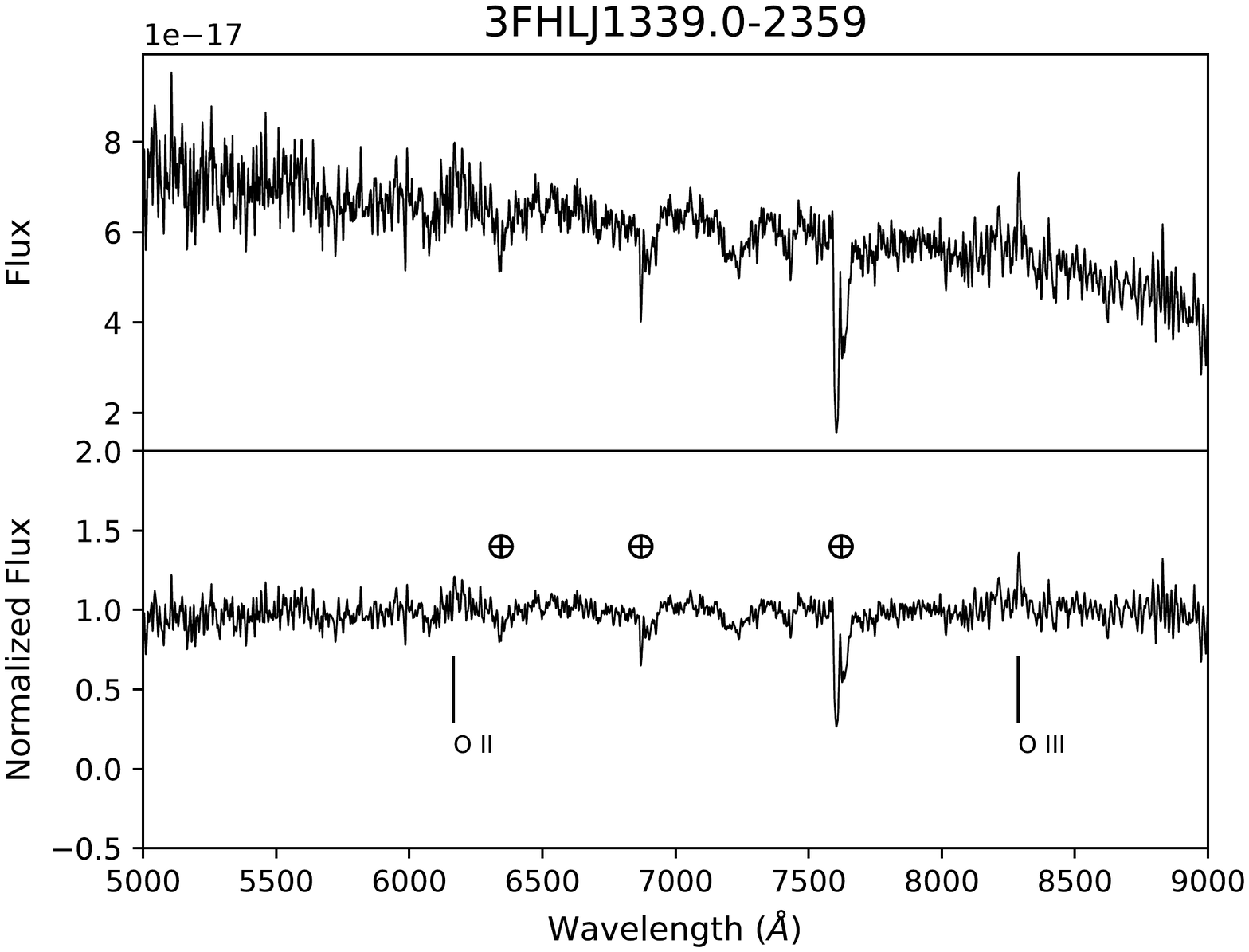}
  \end{minipage}
\begin{minipage}[b]{.5\textwidth}
\vspace{-3.7cm}
  \centering
  \includegraphics[width=0.97\textwidth]{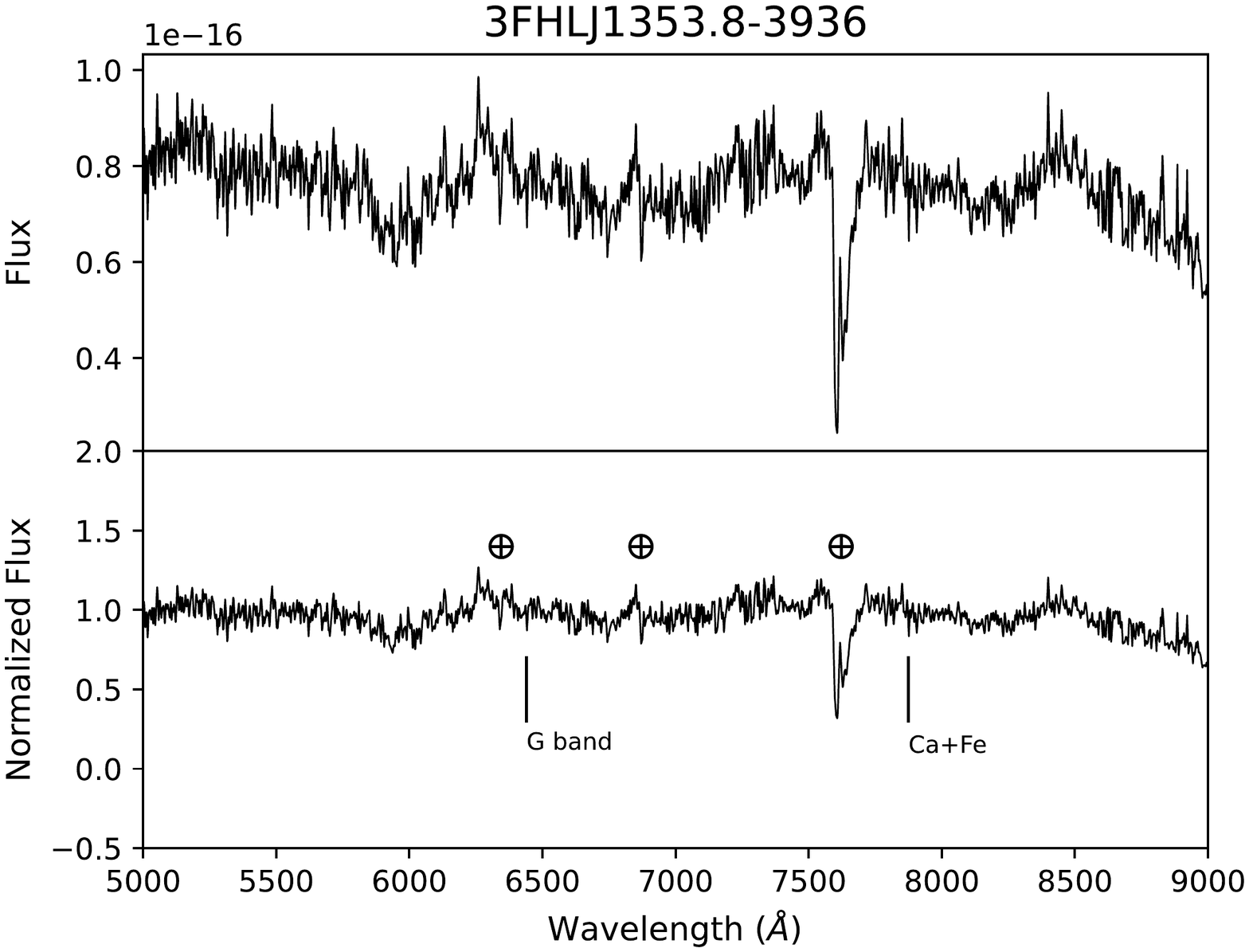}
  \end{minipage}
\end{figure*}

\begin{figure*}
\vspace{-1.7cm}
\begin{minipage}[b]{.5\textwidth}
  \centering
  \includegraphics[width=0.97\textwidth]{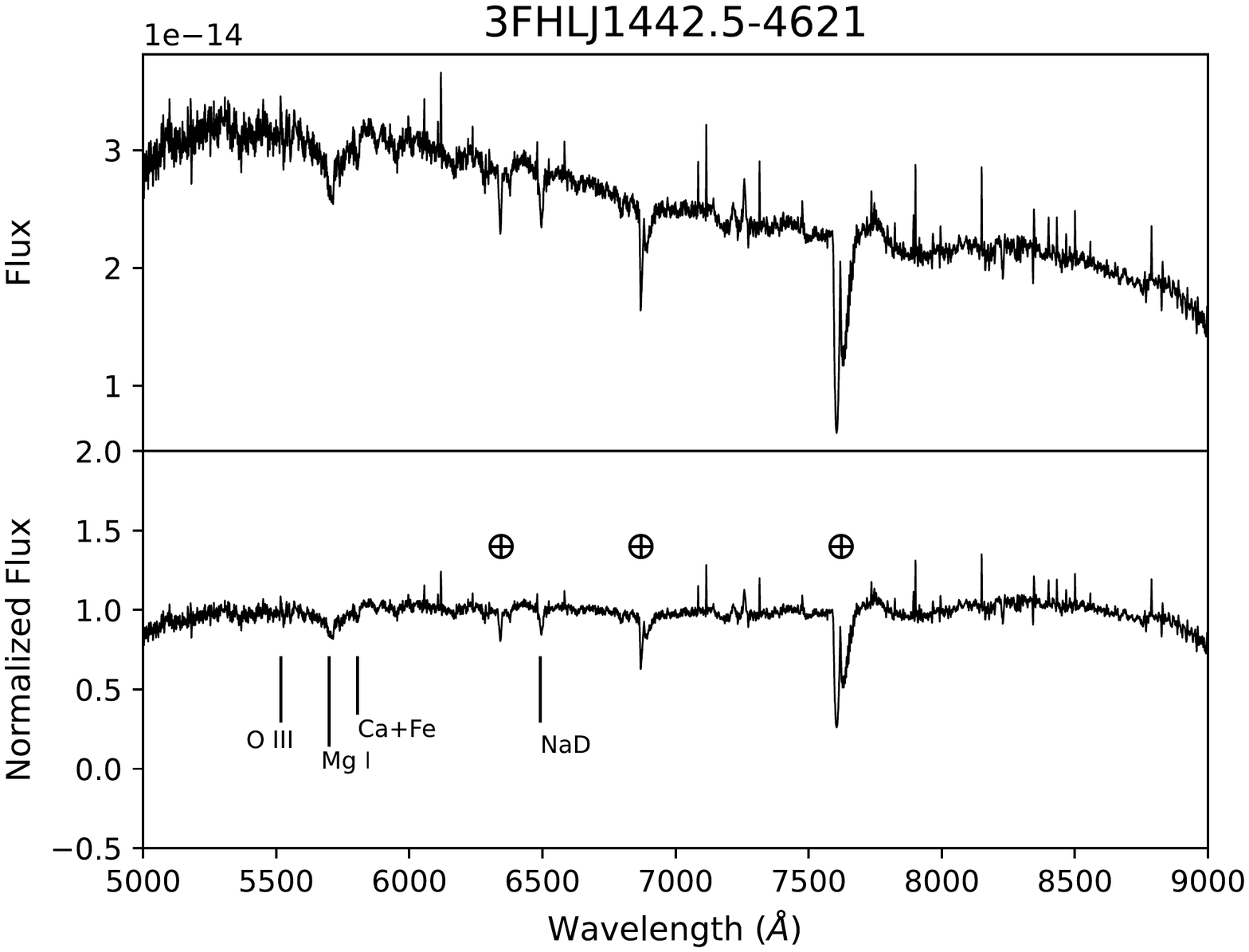}
  \end{minipage}
\begin{minipage}[b]{.5\textwidth}
\vspace{-3.7cm}
  \centering
  \includegraphics[width=0.97\textwidth]{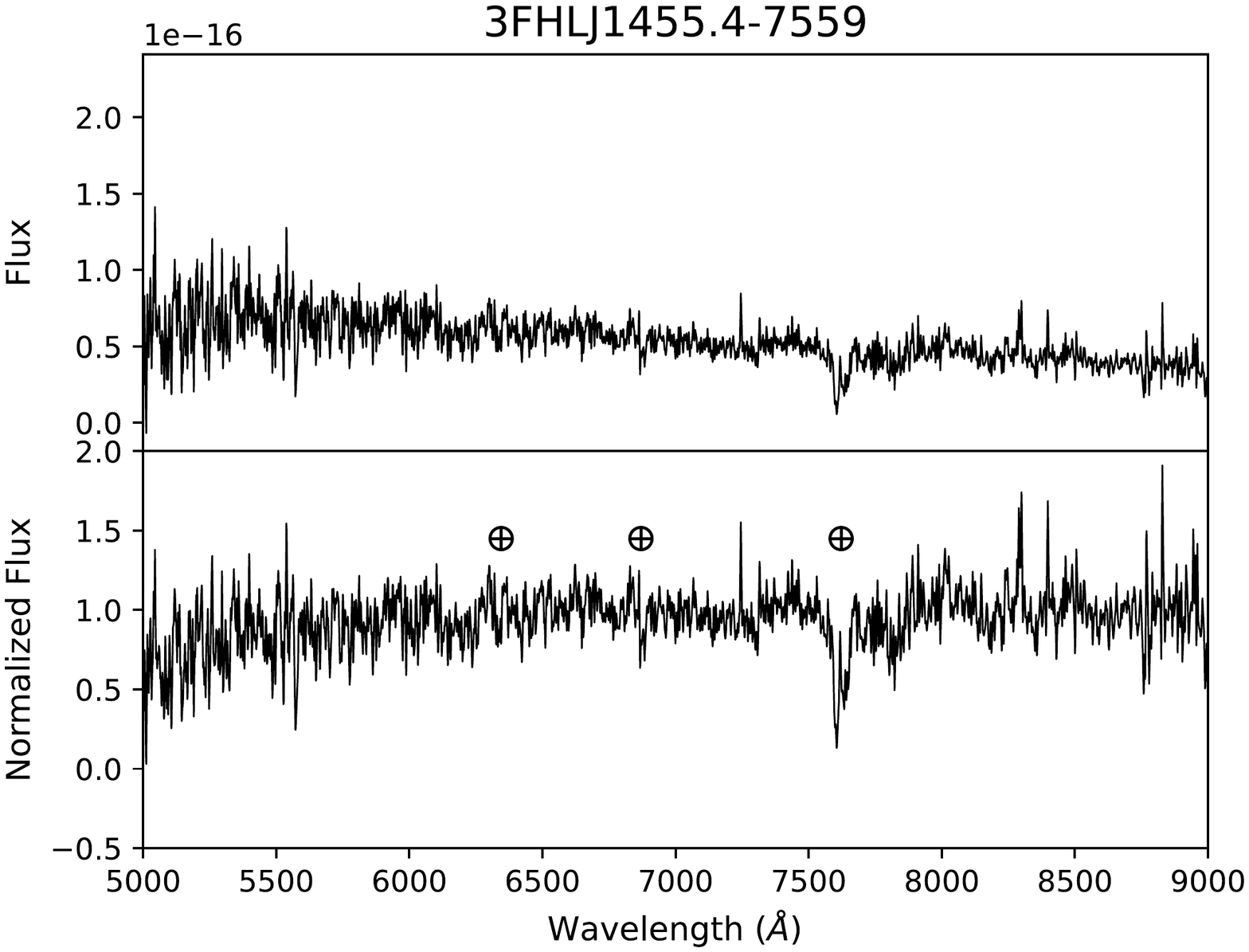}
  \end{minipage}
\begin{minipage}[b]{.5\textwidth}
\vspace{-3.7cm}
  \centering
  \includegraphics[width=0.97\textwidth]{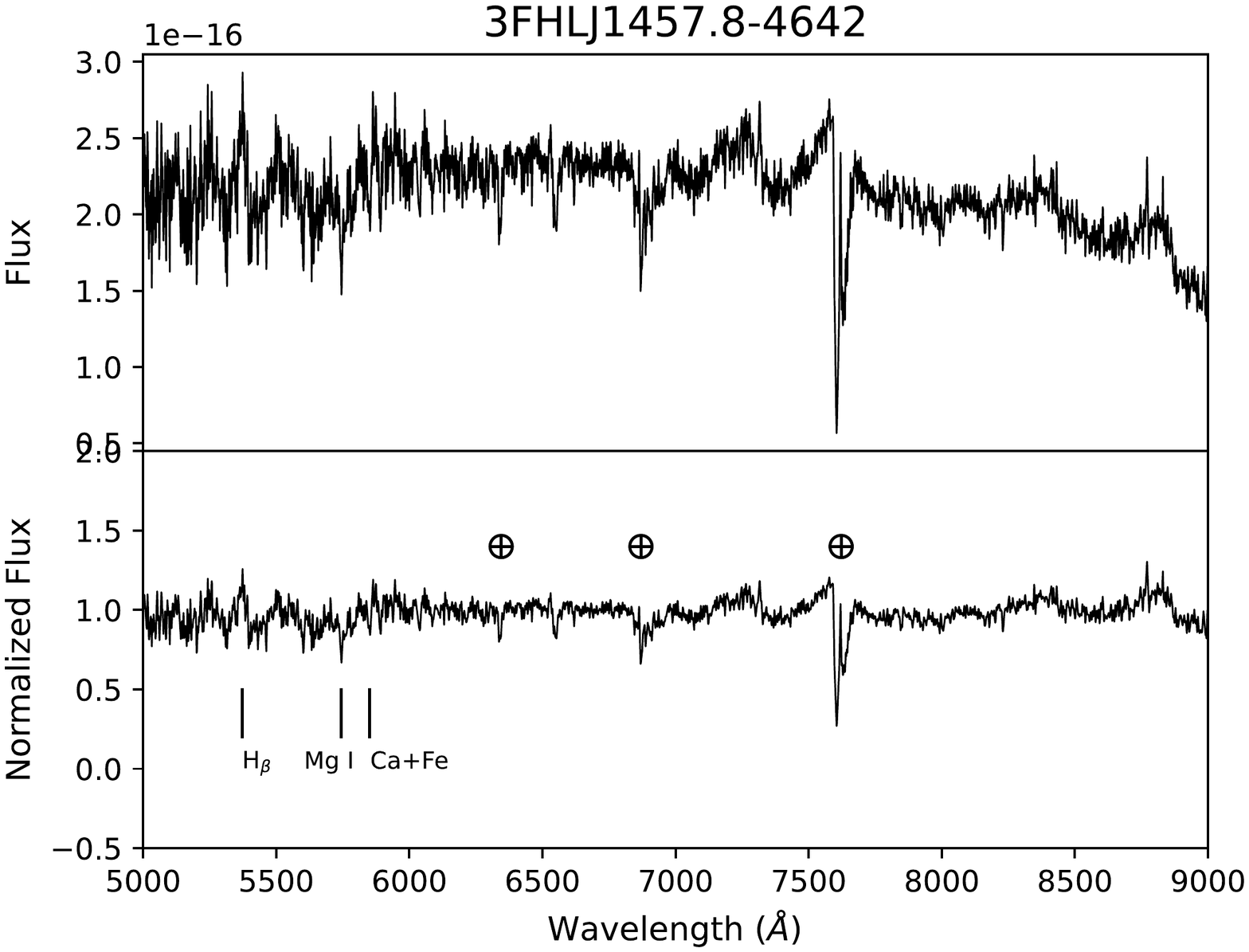}
  \end{minipage}
\begin{minipage}[b]{.5\textwidth}
\vspace{-3.7cm}
  \centering
  \includegraphics[width=0.97\textwidth]{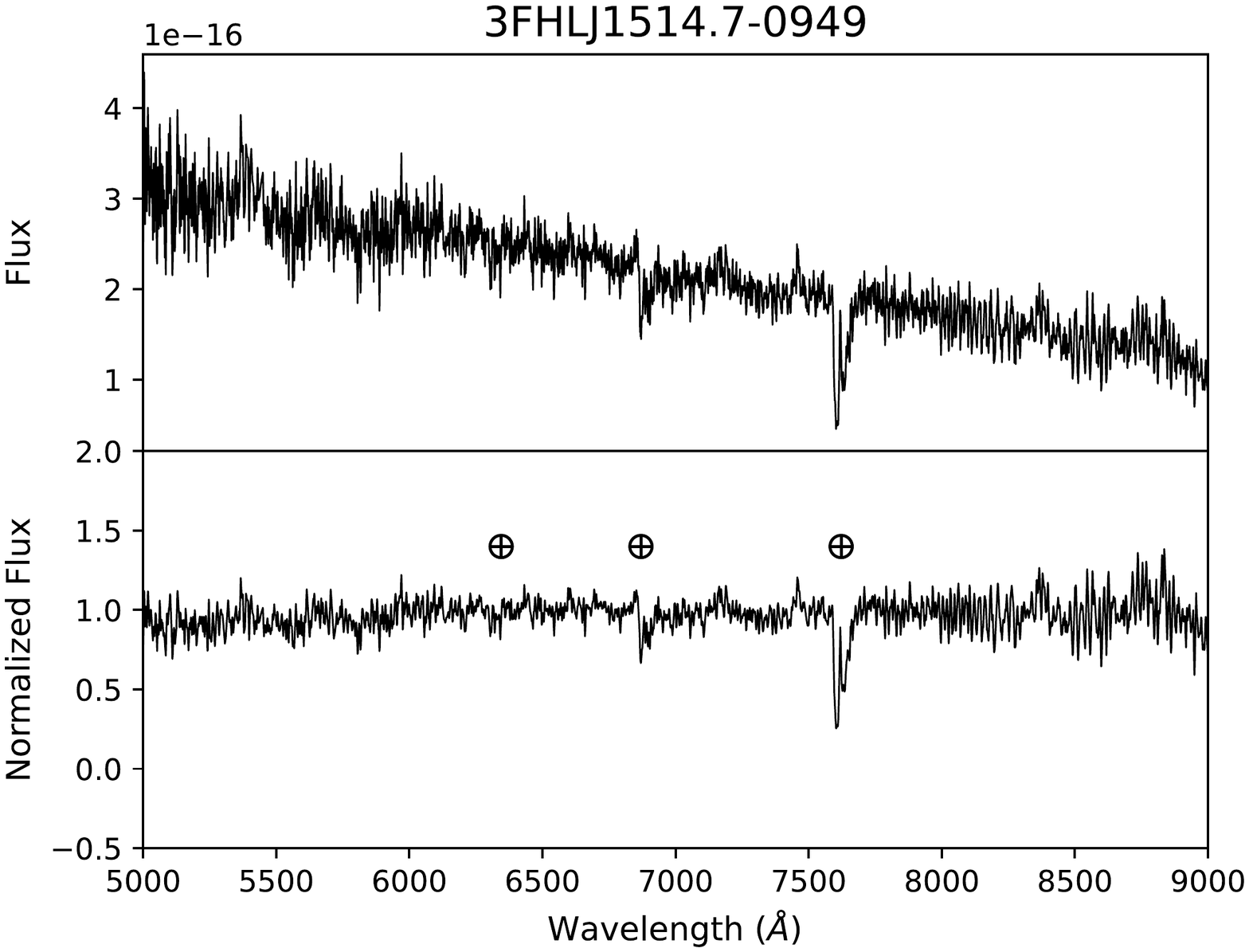}
  \end{minipage}
\begin{minipage}[b]{.5\textwidth}
\vspace{-3.7cm}
  \centering
  \includegraphics[width=0.97\textwidth]{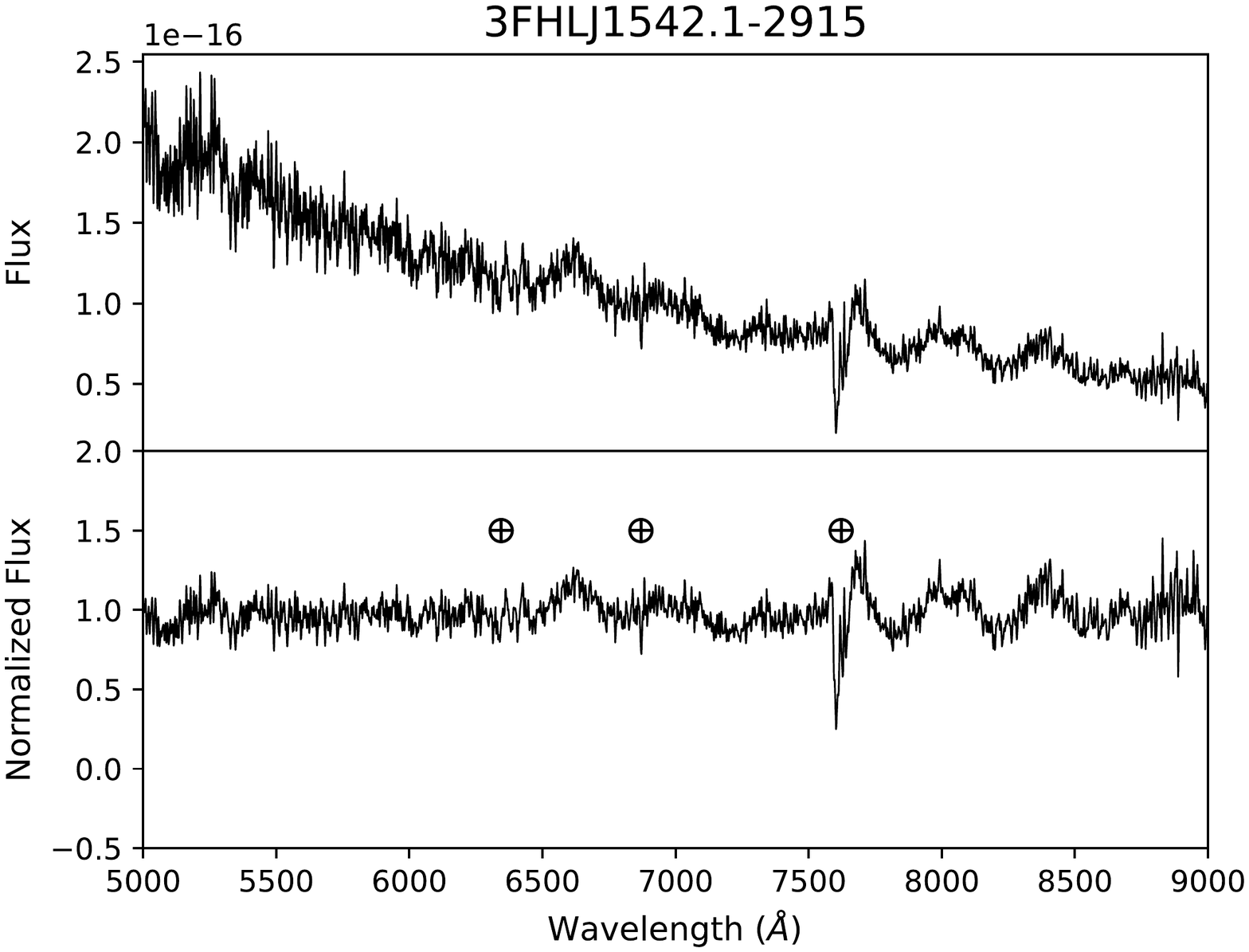}
  \end{minipage}
\begin{minipage}[b]{.5\textwidth}
\vspace{-3.7cm}
  \centering
  \includegraphics[width=0.97\textwidth]{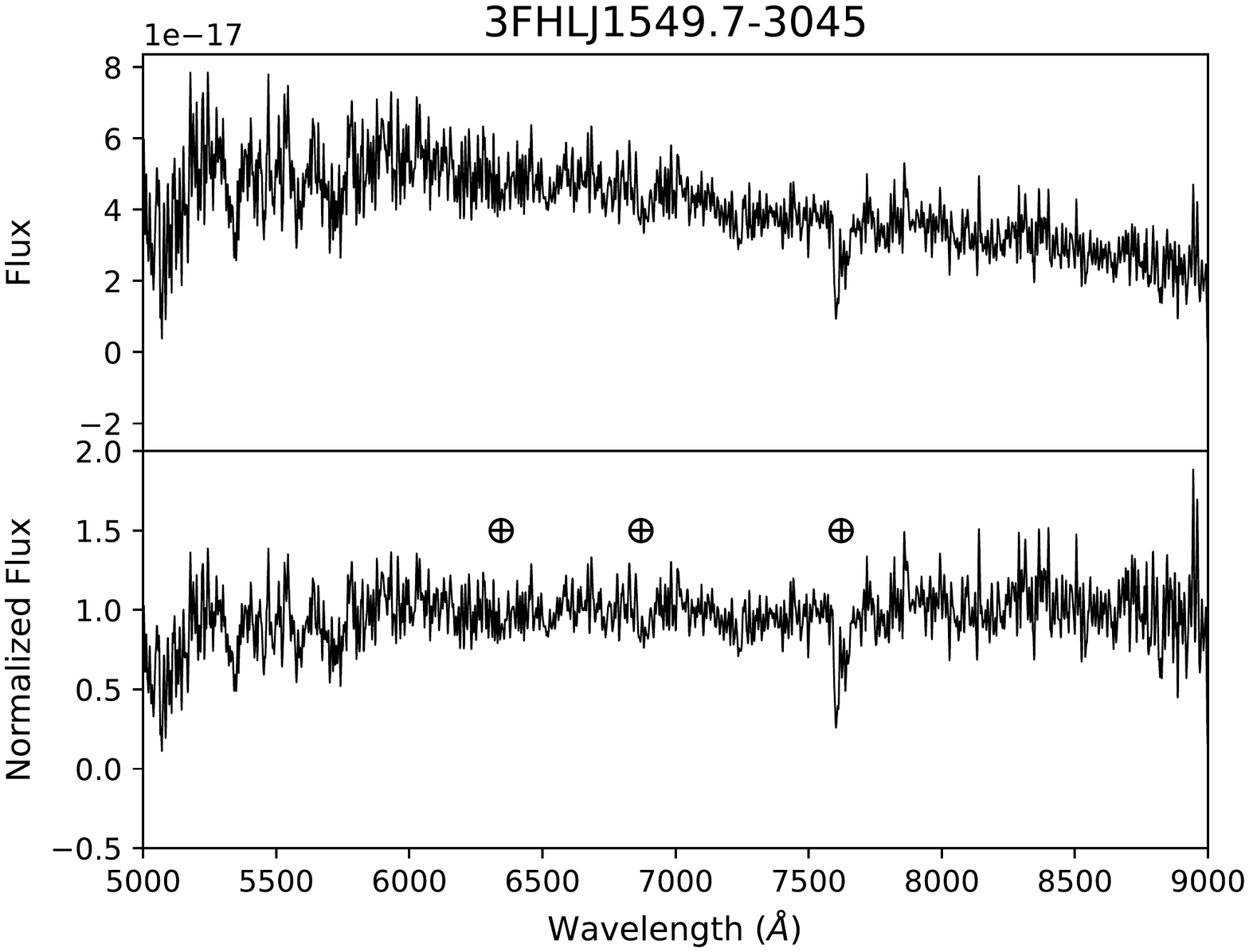}
  \end{minipage}
\end{figure*}

\begin{figure*}
\vspace{-1.7cm}
\begin{minipage}[b]{.5\textwidth}
  \centering
  \includegraphics[width=0.97\textwidth]{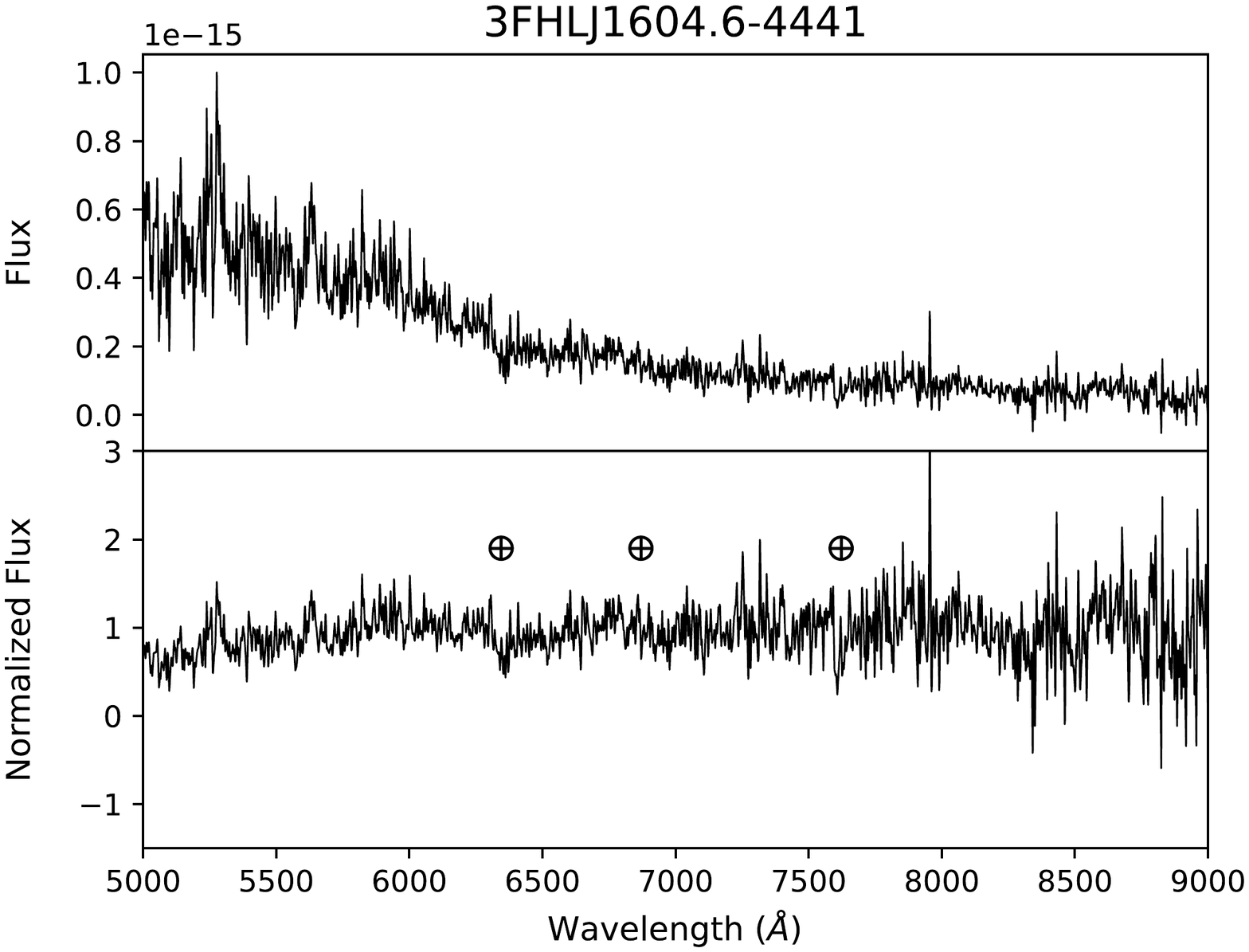}
  \end{minipage}
\begin{minipage}[b]{.5\textwidth}
  \centering
  \includegraphics[width=0.97\textwidth]{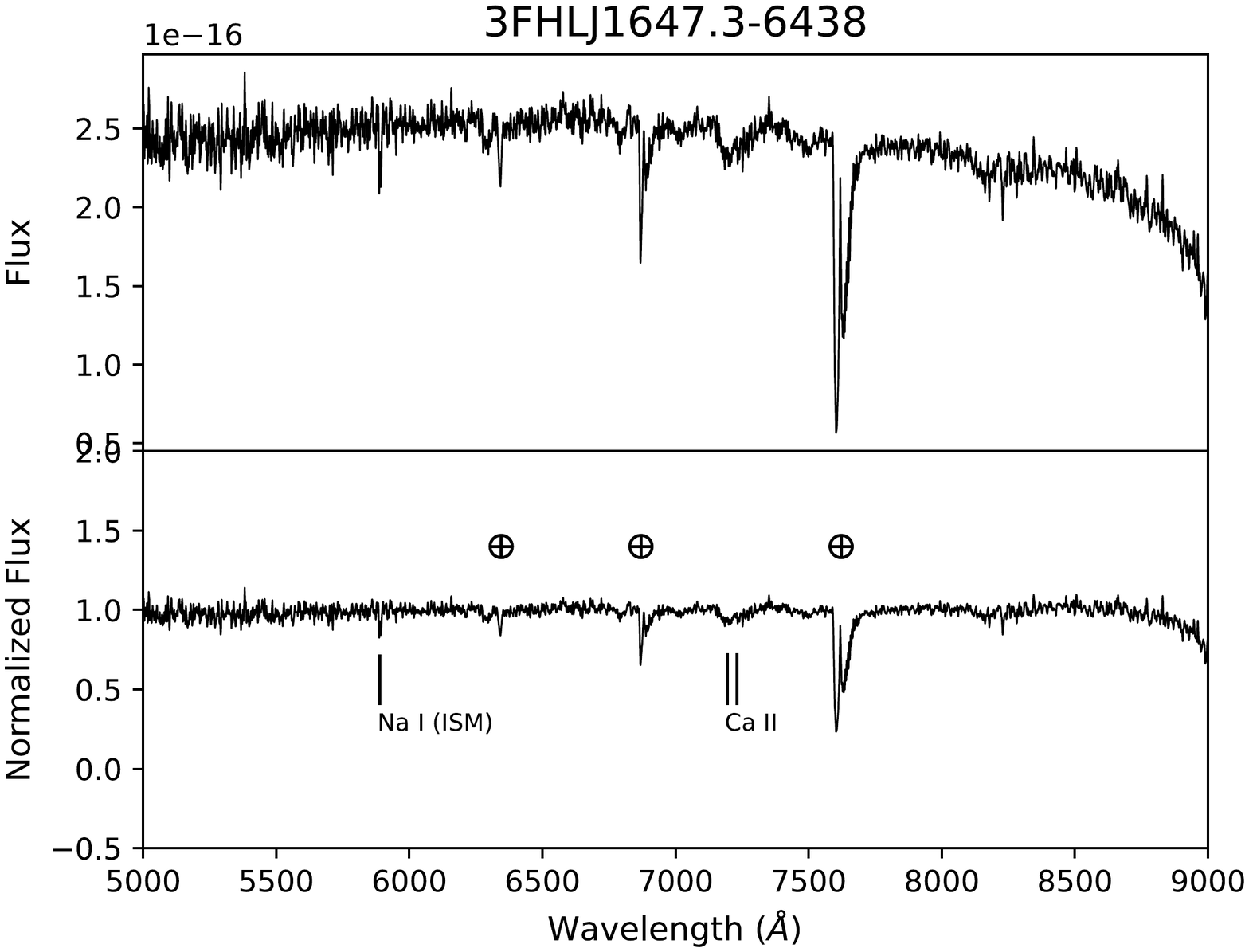}
  \end{minipage}
\label{fig:spec}
\begin{minipage}[b]{.5\textwidth}
\vspace{-3.7cm}
  \centering
  \includegraphics[width=0.97\textwidth]{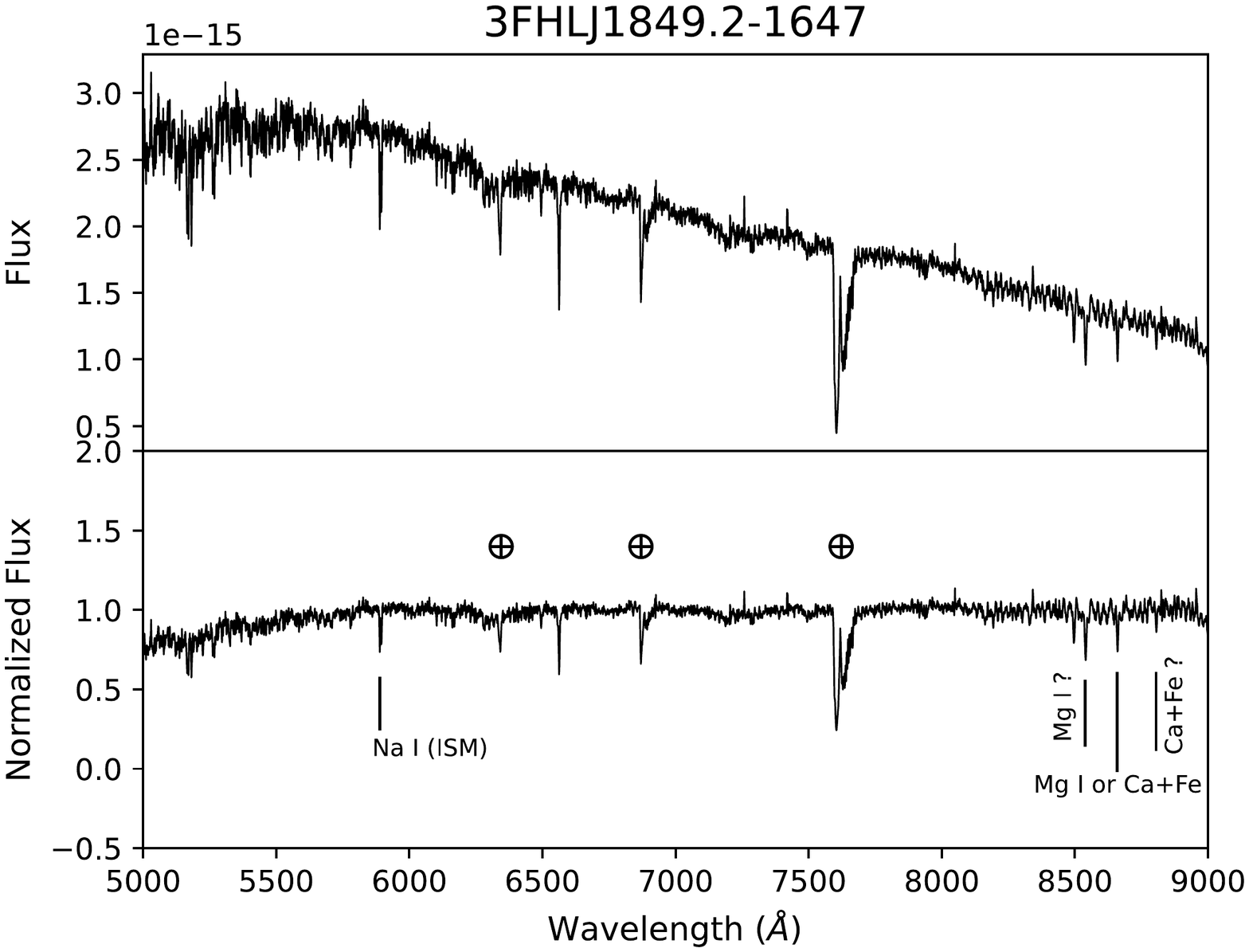}
  \end{minipage}
\begin{minipage}[b]{.5\textwidth}
\vspace{-3.7cm}
  \centering
  \includegraphics[width=0.97\textwidth]{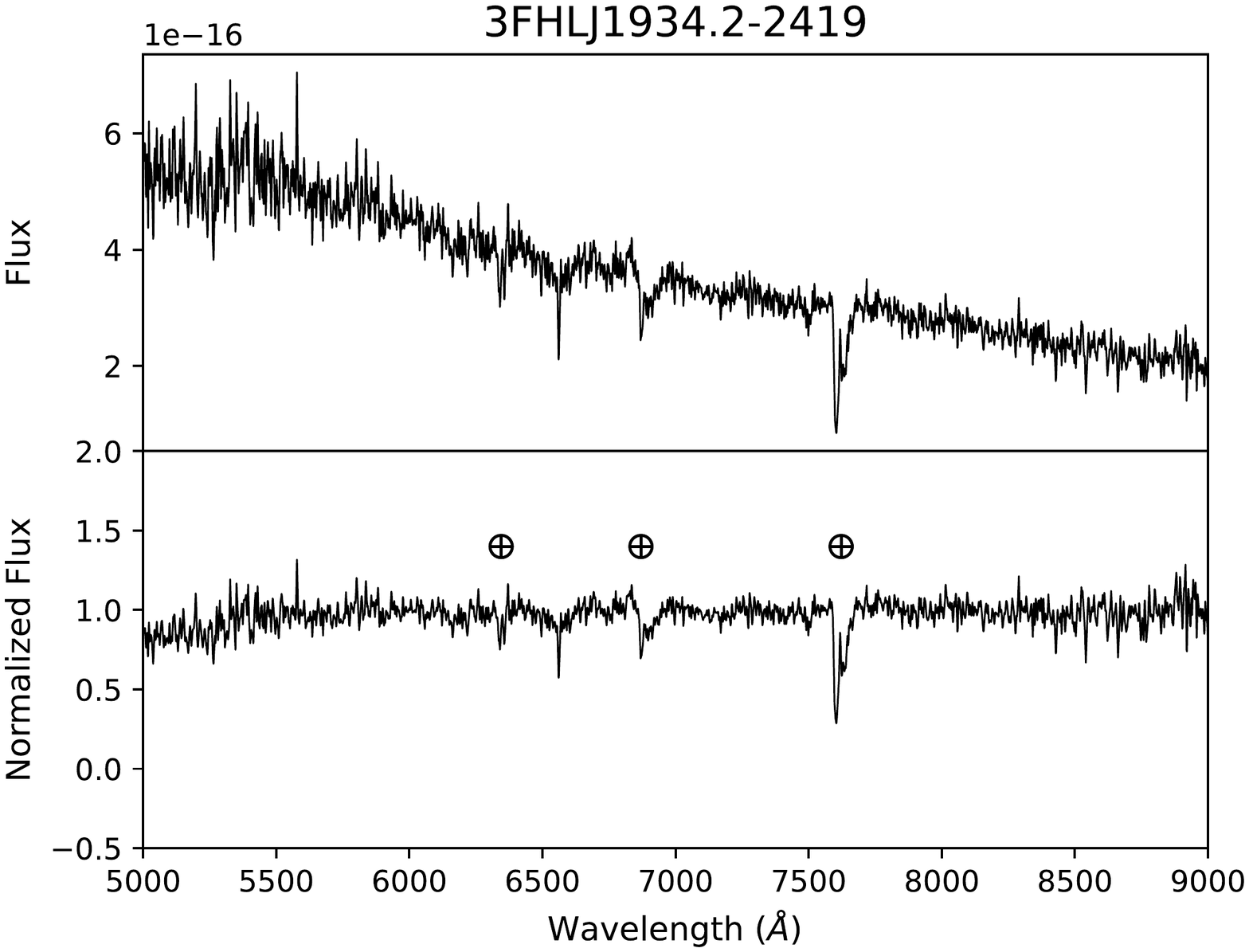}
  \end{minipage}
\begin{minipage}[b]{.5\textwidth}
\vspace{-3.7cm}
  \centering
  \includegraphics[width=0.97\textwidth]{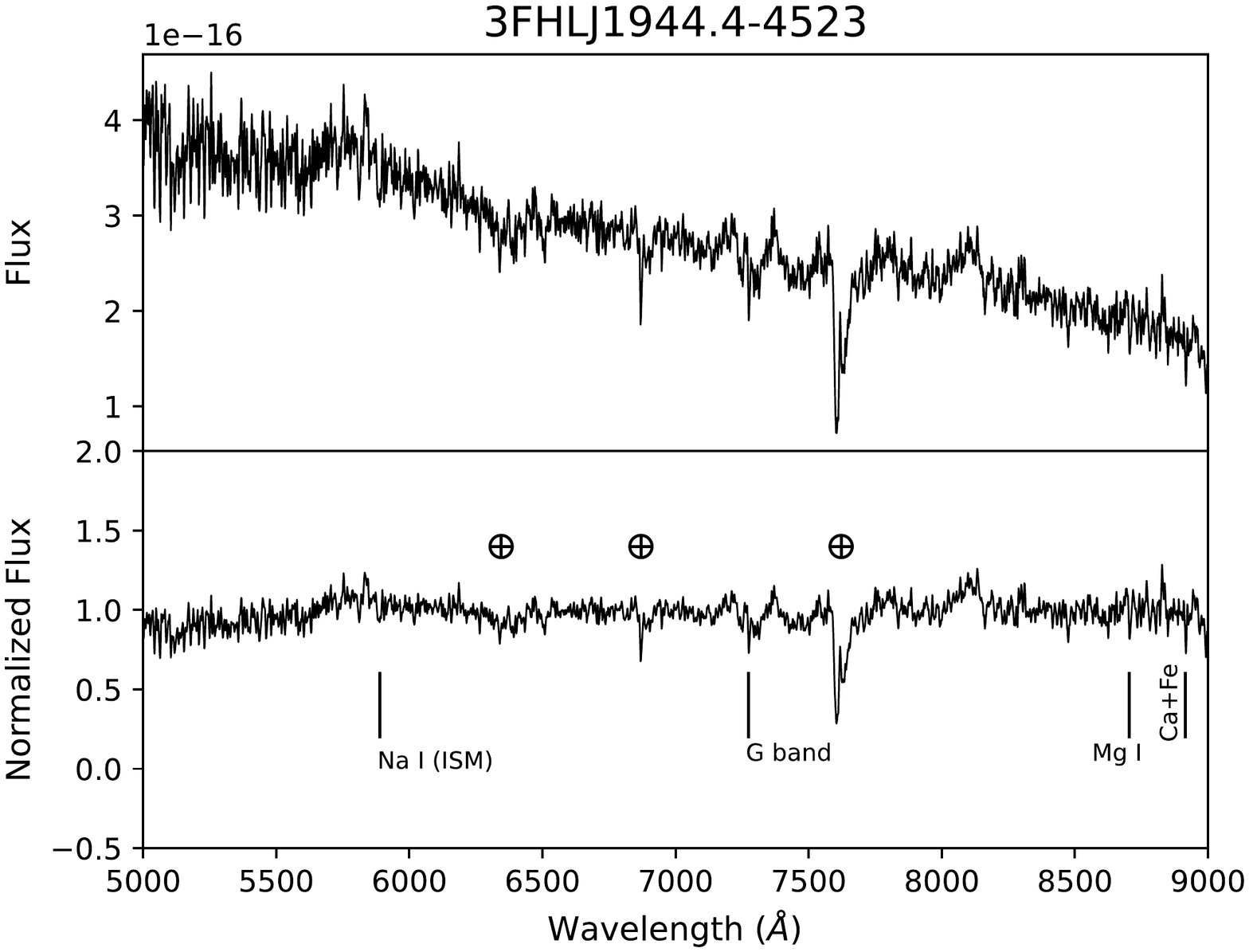}
  \end{minipage}
\begin{minipage}[b]{.5\textwidth}
\vspace{-3.7cm}
  \centering
  \includegraphics[width=0.97\textwidth]{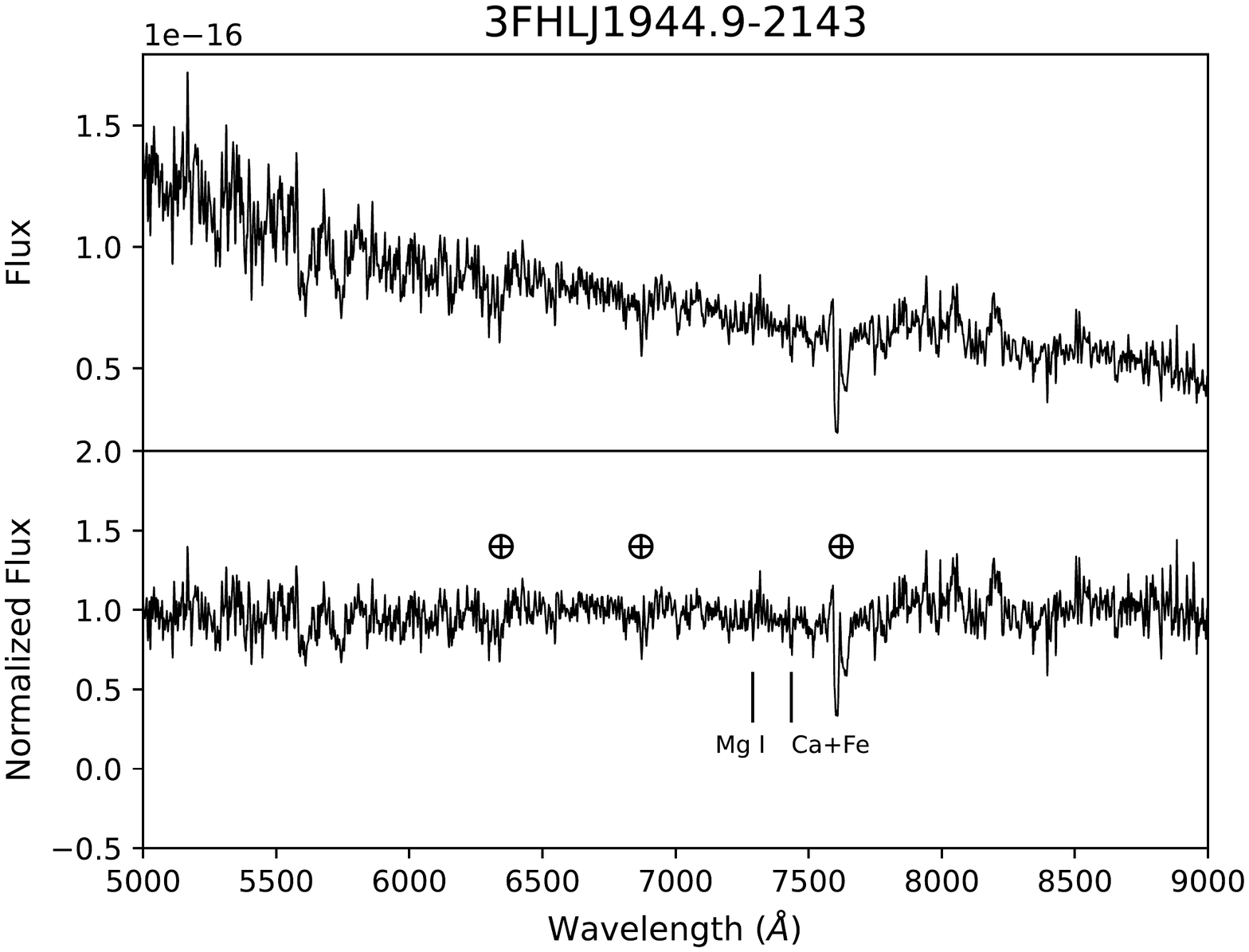}
  \end{minipage}
\end{figure*}

\begin{figure*}
\vspace{-1.7cm}
\begin{minipage}[b]{.5\textwidth}
  \centering
  \includegraphics[width=0.97\textwidth]{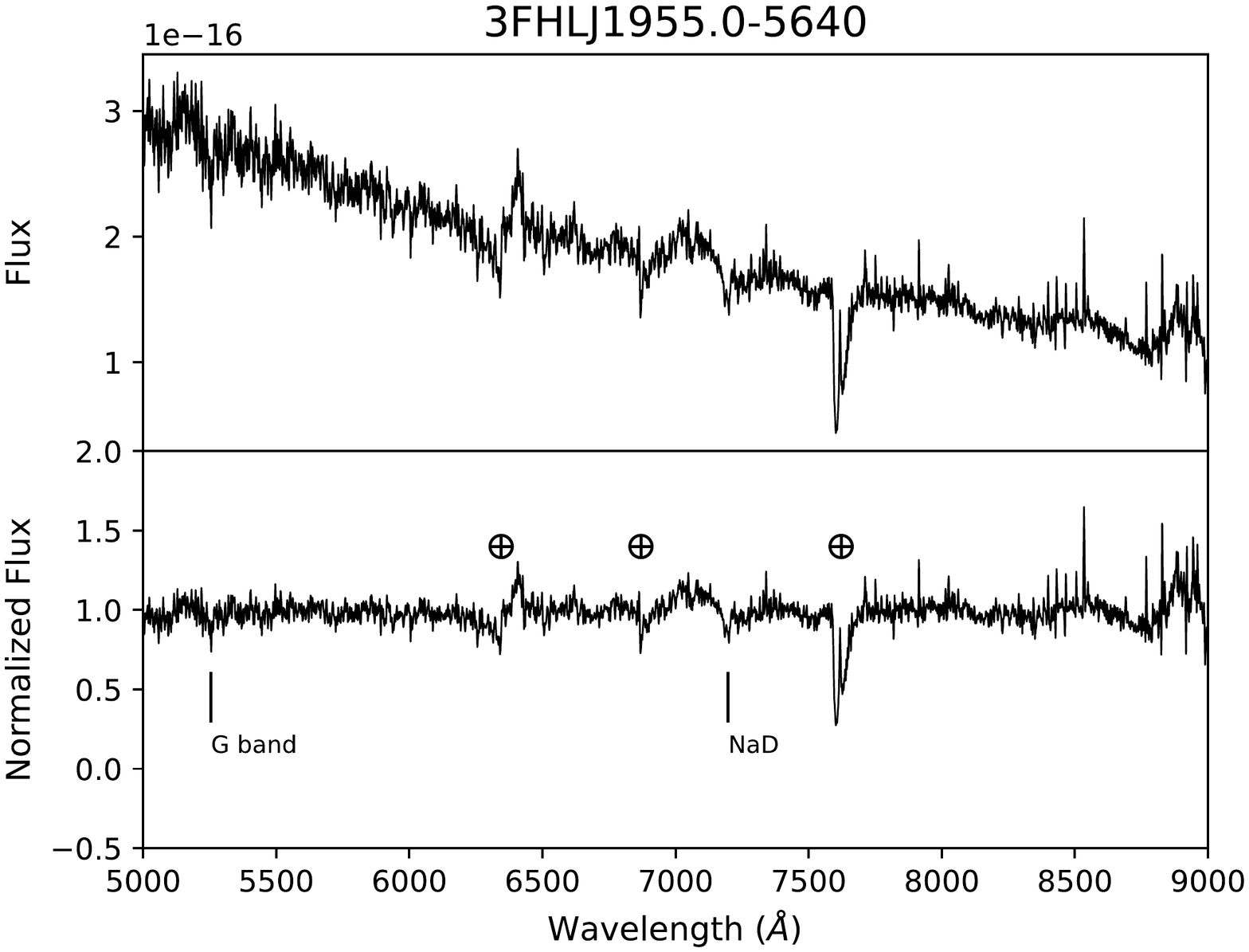}
  \end{minipage}
\begin{minipage}[b]{.5\textwidth}
\vspace{-3.7cm}
  \centering
  \includegraphics[width=0.97\textwidth]{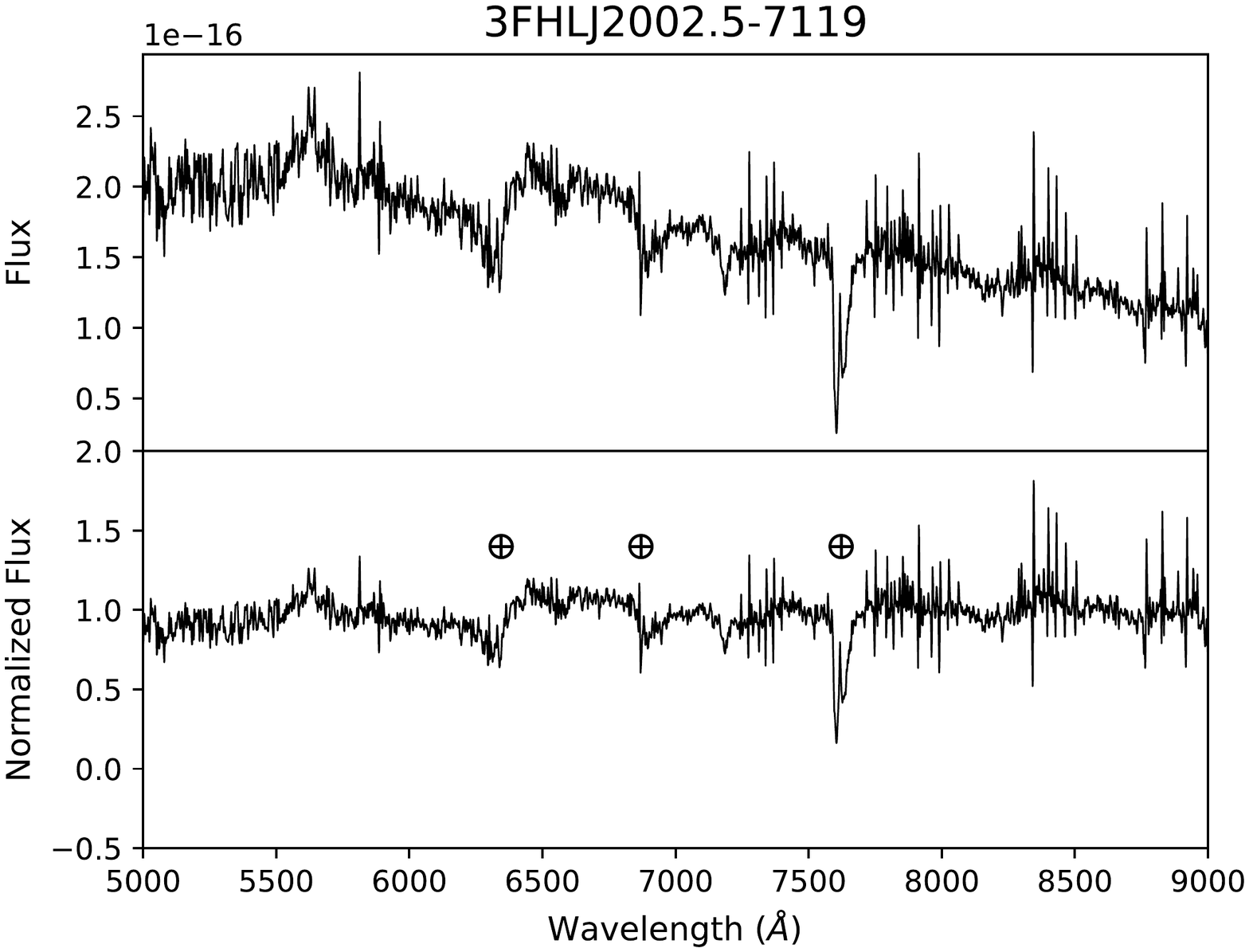}
  \end{minipage}
\begin{minipage}[b]{.5\textwidth}
\vspace{-3.7cm}
  \centering
  \includegraphics[width=0.97\textwidth]{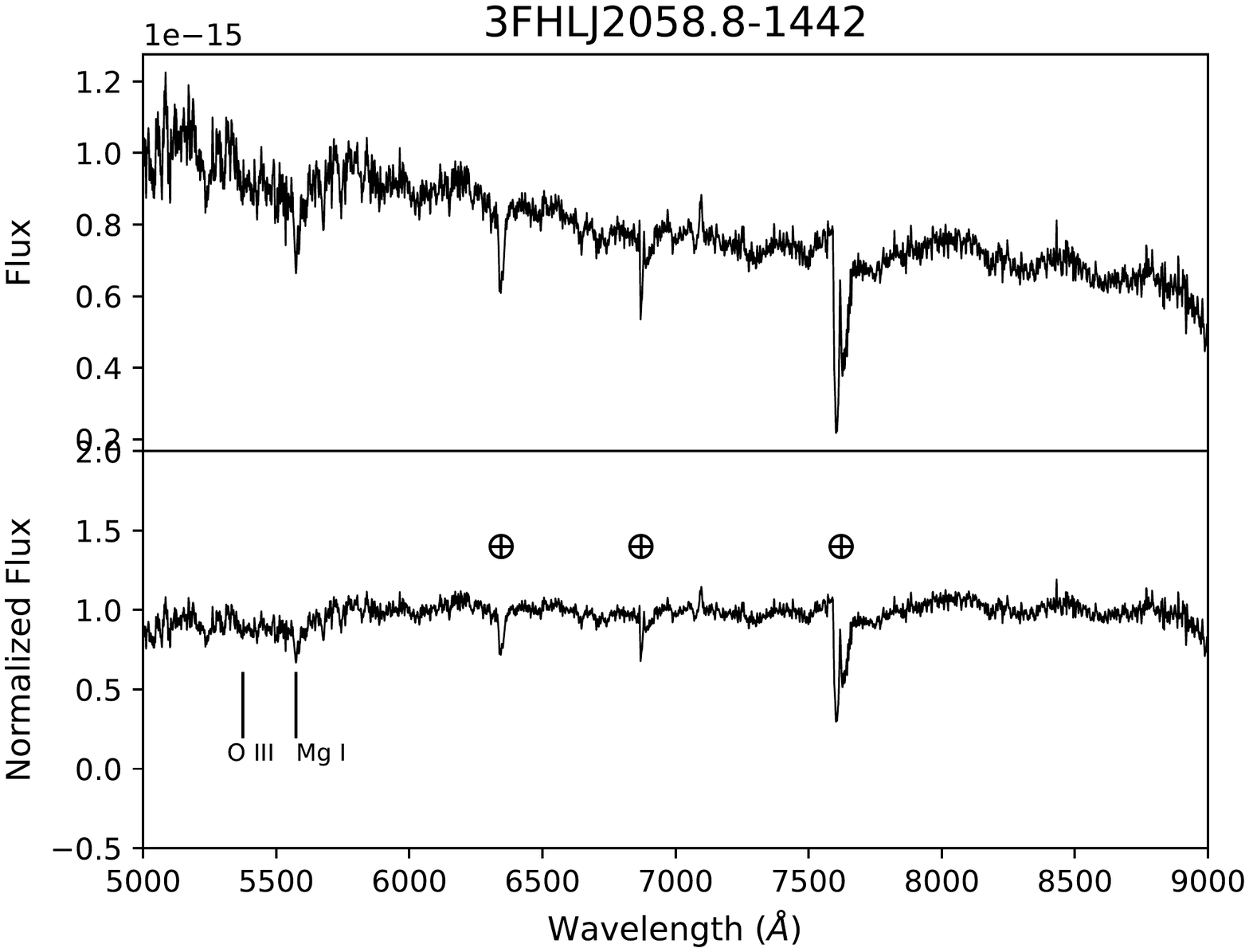}
  \end{minipage}
\begin{minipage}[b]{.5\textwidth}
\vspace{-3.7cm}
  \centering
  \includegraphics[width=0.97\textwidth]{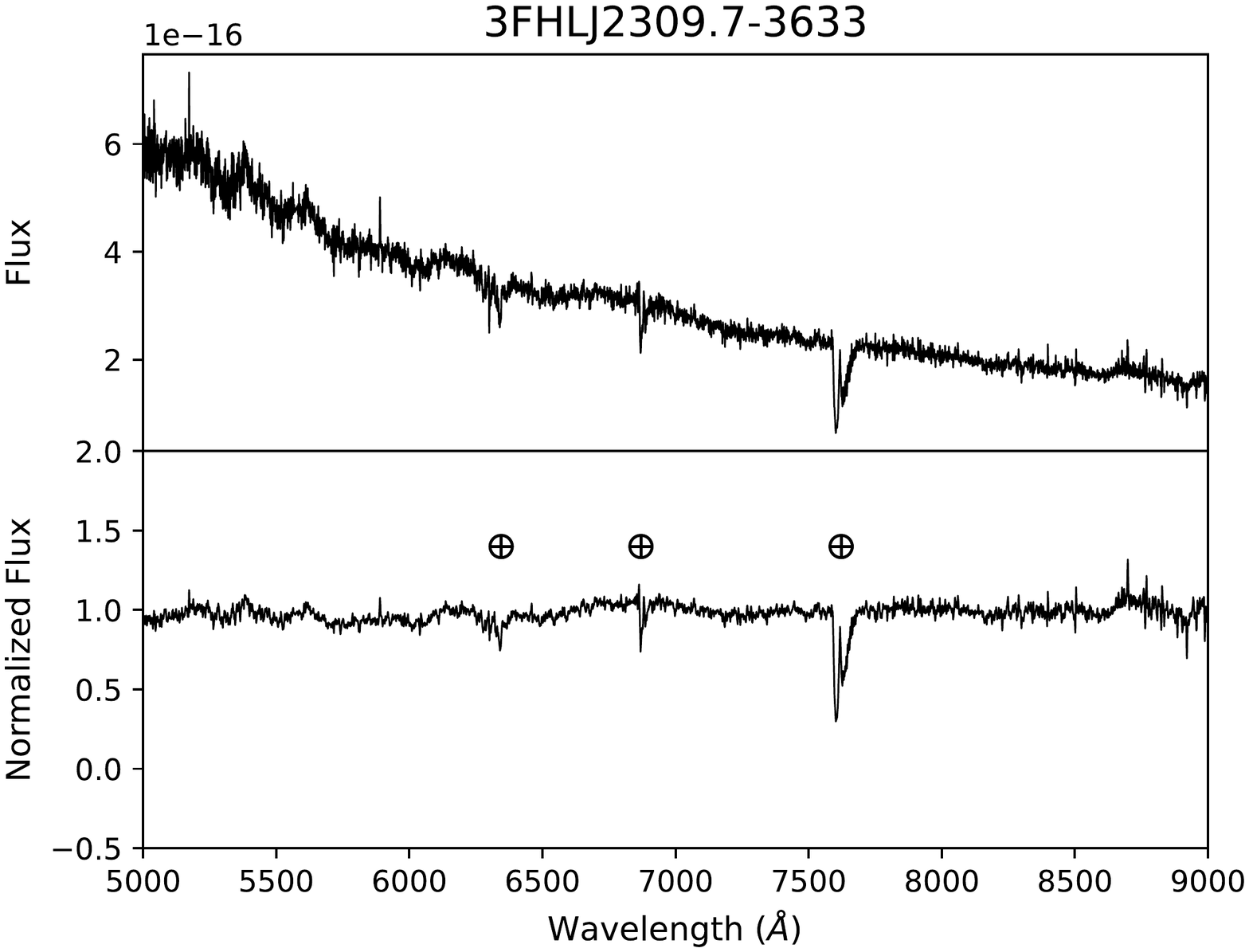}
  \end{minipage}
  \vspace{-2.7cm}
\renewcommand{\thefigure}{1 (Continued)}
\caption{}
\label{fig:spec}
\end{figure*}
\section{Results   \label{results}}
Results obtained from the spectral analysis on the sources have been reported in Table~\ref{tab:redshift} (along with source classifications). Here, we report to the highest precision for which there is agreement amongst different features. The corresponding spectra have been shown in Figure~\ref{fig:spec1}. To make the features more apparent, normalized spectra of our sources are also shown, which were obtained by dividing the flux-calibrated spectrum by a power law fit of the continuum. The signal-to-noise ratio (S/N) of the normalized spectra (reported in Table~\ref{tab:redshift}) were obtained by measuring the ratio in at least five different regions in the featureless part of each spectrum. 

To identify the various emission and absorption features, all spectra were individually inspected. A feature was considered reliable if its presence was verified in all the individual exposures. Sources were classified as FSRQs if the emission lines in their spectra had EW $>$ 5\AA\,, and as BL Lacs if emission lines were observed with EW $<$ 5\AA.

In this work, 25 out of 28 sources have been classified as BL Lacs, with 3 sources being classified as FSRQs. Further, redshifts ($z$) have been provided for 6 sources and lower limits on $z$ for 10 more. The remaining 12 sources exhibited featureless optical spectra.

Comments have been provided for all the sources. Certain source spectra were also observed to be particularly noisy, inhibiting us from identifying any features. Tentative classifications have been provided for such sources.

\subsection{Notes on Individual Sources}
\begin{itemize}
\item {\bf 3FHL J0001.9$-$4155}: The source optical spectrum showed one emission feature at 5450\,\AA\ (EW$=$88\,\AA), resembling those observed in FSRQ spectra. However, the presence of only one such feature prevents us from successfully identifying it; thus, we tentatively classify it as an FSRQ. This source has been reported as an HSP blazar in the fourth catalog of AGN detected by the {\it Fermi}-LAT (4LAC; \citealp{Ajello2020zz}). 

\item {\bf 3FHL J0102.8$-$2001}: The optical spectrum for this source showed an emission feature at 5107 \AA\, (EW=18\,\AA) and an absorption feature at 8084 \AA. These features have been attributed to O {\sc ii} and NaD, respectively. This gives us an estimate of $z=$ 0.37 for the source. This result is in mild disagreement with previously reported value by \cite{1whsp} in the 1WHSP catalog, which found a lower limit of $z>$0.38 for this source. This lower limit on $z$ was obtained by the authors assuming the host to be a standard candle (elliptical galaxy with M$_{R}$=$-$22.9). The non detection of the host allows the authors to set a redshift lower limit for the source. This source has been reported as an HSP blazar in 4LAC.


\item {\bf 3FHL J0659.5$-$6743}: The optical spectrum for this source exhibited no features. This source has been classified as an HSP blazar in 4LAC.

\item {\bf 3FHL J0709.1$-$1525}: The optical spectrum for this source exhibits two absorption features at 5551 \AA\, (H$\beta$) and 6017 \AA\, (Ca$+$Fe), respectively. This result gives us $z>$ 0.142 for this source which we classify as a BL Lac. This result is in mild disagreement with the redshift obtained by \cite{paiano2020} of $z=0.142$. Apart from the absorption features, the authors also found an emission feature at 7518 \AA\, that they attributed to N {\sc ii} (6584 \AA). This feature was not observed in our spectra for this source. \cite{paiano2020} obtained the optical spectra for this source with the OSIRIS spectrograph at the 10.4\,m GTC at Roque de Los Muchachos telescope facility. This source is an HSP blazar in 4LAC.

\item {\bf 3FHL J0737.5$-$8247}: The optical spectrum for this source displayed no features. This source has no reported SED class in the \textit{Fermi}-LAT catalogs.

\item {\bf 3FHL J0953.3$-$7659}: The optical spectrum for this source showed no features and has been classified as an HSP blazar in 4LAC.

\item {\bf 3FHL J1042.2$-$4128}: {In the optical spectrum for this source, features identified were; G$-$band absorption feature at 5210 \AA\, and Mg {\sc i} feature at 6274 \AA, respectively.} These two features allowed us to establish a lower limit redshift of $z>$ 0.21 for this source. A featureless optical spectrum for this source was obtained by \cite{menezes2020z} using the 4.1\,m SOAR telescope at Cerro Pachón, Chile. 
We classify this source as a BL Lac. This source has been reported as an HSP blazar in 4LAC.

\item {\bf 3FHL J1130.7$-$3137}: The optical spectrum of this source displayed two emission features at 5609 \AA\, and 5770 \AA, corresponding to H$\beta$ (EW = 2.4 \AA) and O {\sc iii} (EW = 1.1 \AA), respectively. This yields a redshift of $z=$ 0.15 for the source. The 6dF Galaxy Survey \citep{6df} obtained a spectrum for this source (obtained by the 6dF fibre-fed multi-object spectrograph at the United Kingdom Schmidt Telescope) from which they derived a redshift of 0.15, and classified the source as a BL Lac, in agreement with our results. This source has been classified as an HSP blazar in 4LAC.

\item {\bf 3FHL J1225.4$-$3447}: In the optical spectrum for this source, two absorption features were detected at 5994 \AA\,, corresponding to G$-$band, and 7336 \AA, corresponding to Ca$+$Fe lines, respectively. This produces a redshift of $z>$0.39 for the source, which we classify as a BL Lac. This source has been reported as an HSP blazar in 4LAC.  

\item {\bf 3FHL J1328.9$-$5607}: The optical spectrum obtained for this source was noisy, but we were able to identify two absorption features at 5063 \AA\, and 6099 \AA, ascribed to G$-$band and Mg {\sc i} lines, respectively. This yields a lower limit of $z>$ 0.17 for this source. Our proposed lower limit value is compatible with the value reported by \cite{Shaw_2013} of $z>$ 0.13. This lower limit on $z$ was obtained by the authors assuming the host to be a standard candle (elliptical galaxy with M$_{R}$=$-$22.5).
While the authors were unable to  determine  whether  the  source  is  a  BL  Lac  or an FSRQ, they reported it as an LSP blazar, in agreement with 4LAC. We classify this source as a BL Lac.

\item {\bf 3FHL J1339.0$-$2359}: This source has been reported as a BL Lac in the 4FGL catalog with a $z$ of 0.657. Possible emission features at 6165 \AA \, (O {\sc ii}, EW = 47 \AA) and 8287 \AA \, (O {\sc iii}, EW = 22 \AA) agree with the measured redshift of $z=$0.65. We classify this source as an FSRQ. The redshift for this source was first measured by \cite{ocars}, in the Optical Characteristic of Astrometric Radio Sources (OCARS) catalog, which was then referred by the VLBI catalog by \cite{vlbi2017}.  This source has been reported as an LSP blazar in 4LAC. 

\item {\bf 3FHL J1353.8$-$3936}: Our fairly noisy optical spectrum for this source allowed us to find two absorption features at 6440 \AA\, (G$-$band) and 7875 \AA\, (Ca$+$Fe), respectively, yielding a redshift of $z>$0.49. We classify this source as a BL Lac. This source has been reported as an HSP blazar in 4LAC.

\item {\bf 3FHL J1442.5$-$4621}: This source displays absorption lines at 5698 \AA, 5806 \AA, and 6492 \AA, which are attributed to Mg {\sc i}, Ca$+$Fe, and NaD lines, respectively. An emission line at 5517 \AA\ (O {\sc iii}, EW = 1.5\AA) was also observed. This leads to a redshift of $z=$ 0.102, confirming the measurement by the 6dF Galaxy Survey \citep{6df}. This source has been classified as a BL Lac in the 4FGL catalog. Our results are in agreement with this classification. This source has been reported as an HSP blazar in 4LAC.

\item {\bf 3FHL J1455.4$-$7559}: This source displayed a featureless optical spectrum and has been reported by the 4LAC to be an HSP blazar.

\item {\bf 3FHL J1457.8$-$4642}: In the optical spectrum, absorption lines were detected at 5744 \AA\, and 5851 \AA. These lines are attributed to Mg {\sc i} and Ca$+$Fe lines, respectively. An emission line detected at 5372 \AA\, (EW = 0.6 \AA) could possibly be due to H$\beta$. These features give us a redshift measurement of $z=$ 0.10. We classify this source as a BL Lac. This source has no reported SED class in the {\it Fermi}-LAT catalogs.

\item {\bf 3FHL J1514.7$-$0949}: This source displayed a featureless optical spectrum and has been classified by the 4LAC to be an LSP blazar.

\item {\bf 3FHL J1542.1$-$2915}: A particularly noisy spectrum was obtained for this source, due to which we were unable to identify any features. Thus, we tentatively classify this source as a BL Lac. This source is an HSP blazar in 4LAC.

\item {\bf 3FHL J1549.7$-$3045}: The optical spectrum of this source was featureless. This source is an HSP blazar in 4LAC.

\item {\bf 3FHL J1604.6$-$4441}: This source's optical spectrum was observed to be featureless. This object has been classified as an LSP blazar in 4LAC.

\item {\bf 3FHL J1647.3$-$6438}: The optical spectrum for this source exhibits an absorption doublet at 7195 \AA\ and, 7230 \AA\,. These lines correspond to Ca {\sc ii} lines, yielding a lower limit of $z>$ 0.82. We classify this source as a BL Lac. This source is classified as an LSP blazar in 4LAC.

\item {\bf 3FHL J1849.2$-$1647}: The optical spectrum for this source displayed three absorption features at \textbf{8537} \AA, 8659 \AA, and 8805 \AA. If the first two features are attributed to Mg {\sc i} and Ca$+$Fe, then we obtain $z>$0.64. On the other hand, if the feature at 8659 \AA\ is due to Mg {\sc i}, then the feature at 8805 \AA\, can be attributed to Ca$+$Fe, yielding $z>$0.67. We classify this source as a BL Lac. This source is classified as an HSP blazar in 4LAC.

\item {\bf 3FHL J1934.2$-$2419}: This source displayed a featureless optical spectrum and has been reported by the 4LAC catalog to be as HSP blazar.

\item {\bf 3FHL J1944.4$-$4523}: In the optical spectrum for this source, three absorption features were detected. These three features are  at 7274 \AA\, (G$-$band), 8704 \AA\, (Mg {\sc i}), and 8916 \AA\, (Ca$+$Fe), yielding $z>$0.68. We classify this source as a BL Lac. This source has no reported SED class in the {\it Fermi}-LAT catalogs.

\item {\bf 3FHL J1944.9$-$2143}: In this optical spectrum, absorption features were detected at 7289 \AA\, and 7435 \AA. These correspond to Mg {\sc i} and Ca$+$Fe lines, respectively, yielding  a lower limit, $z>$ 0.41. This result agrees with the value reported by \cite{2whsp} in the 2WHSP catalog. This lower limit value was obtained by the authors using the same technique as the 1WHSP catalog (described for the source 3FHL J0102.8--2001). We classify this source as a BL Lac. This source has no reported SED class in the {\it Fermi}-LAT catalogs.

\item {\bf 3FHL J1955.0$-$5640}: In the optical spectrum for this source, two absorption features were identified at 5255 \AA\, (G$-$band) and 7198 \AA\, (NaD). This gives us a lower limit of $z>$ 0.22 for this source. We classify this source as a BL Lac. Our results are in mild disagreement with the redshift value reported by \cite{marchesini2019} of $z=$ 0.22. The authors also observed only absorption features for this source that they attributed to the host galaxy instead. Their optical spectra for this source was obtained using the 4.1\,m Southern Astrophysical Research Telescope (SOAR), located in Chile. This source is an HSP blazar according to 4LAC.

\item {\bf 3FHL J2002.5$-$7119}: This source exhibited a featureless optical spectrum and has no reported SED class in the 4LAC.

\item {\bf 3FHL J2058.8$-$1442}: This source displayed an emission peak at 5375 \AA\, that is due to O {\sc iii} (EW = 0.4 \AA) and an absorption feature at 5574 \AA, due to Mg {\sc i}. This allows us to confirm the redshift of $z=$ 0.07 for this source as reported by \cite{6df}. We classify this source as a BL Lac. This source has no reported SED class in the {\it Fermi}-LAT catalogs.

\item {\bf 3FHL J2309.7$-$3633}: The source exhibited a featureless optical spectrum and has been reported as an HSP blazar in the 4LAC.

\end{itemize}
\hspace{-1cm}
\begin{figure*}
\begin{minipage}[b]{.5\textwidth}
  \centering
  \includegraphics[width=1.02\columnwidth]{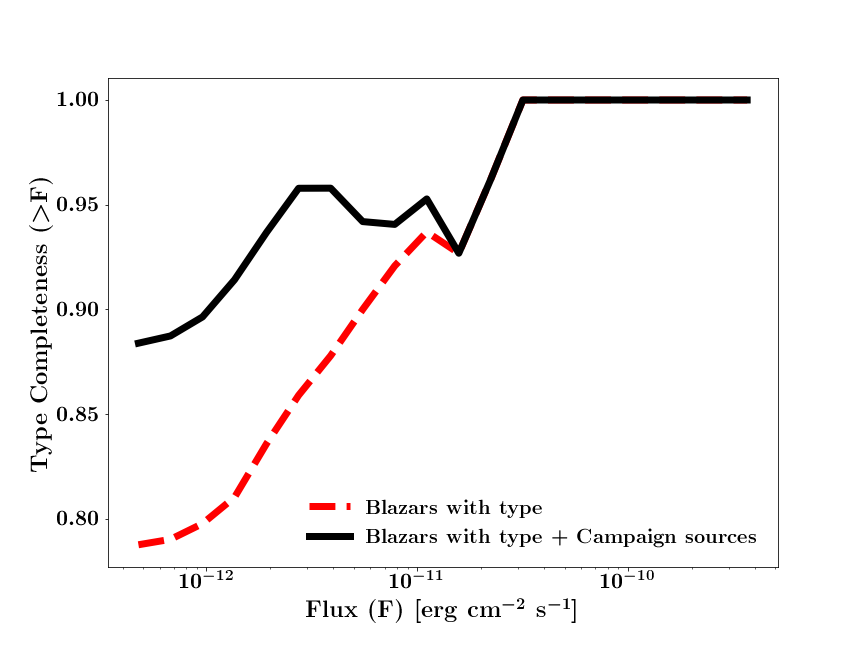}
  \end{minipage}
  \hspace{-0.3cm}
\begin{minipage}[b]{.5\textwidth}
  \centering
  \includegraphics[width=1.03\columnwidth]{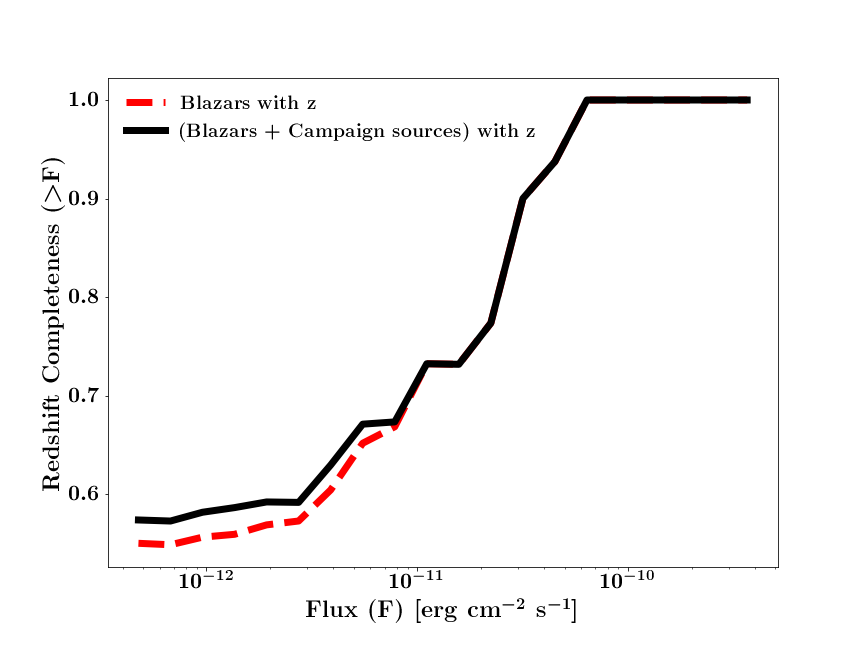}
  \end{minipage}
\caption{Completeness as a function of energy flux in the 3FHL energy range (10 GeV -- 2 TeV). At each flux, f$>$F, we take the total number of sources with type/redshift and divide it by the total number of sources. 
\textit{Left panel:} Type completeness has been computed by considering the number of blazars with classification (or type) with respect to the total number of blazars (red dashed line). Including the contribution of this campaign to the number of classified blazars (solid black line) shows an increase in the overall completeness. \textit{Right panel:} Redshift completeness for blazars with a measured $z$ value with respect to the total number of blazars has been plotted in dashed red line. After including the contribution of this campaign to the overall number of blazars with $z$ (solid black line), there is an increase in the $z$ completeness.}
\label{fig:compl}
\end{figure*}
\section{Discussion and Conclusions}\label{conc}
This identification campaign to classify BCUs in the 3FHL catalog and to measure their redshift began in 2017, when 36 unassociated sources (out of 52 chosen) from 3FHL catalog were classified as BL Lacs with the help of machine-learning algorithms \citep{Kaur18}. For the second part of this campaign, 28 sources were observed with the 4\,m telescope at the Kitt Peak National Observatory (KPNO) in Arizona, in order to perform optical spectroscopy. These results were published in \cite{marchesi18}, where 27 sources were classified as BL Lacs and one as an FSRQ. Redshift measurements or lower limits were also provided for 7 of these sources. The third part of the campaign, for which spectra for 23 sources were obtained using the 4\,m telescope at Cerro Tololo Inter-American Observatory (CTIO) in Chile, resulted in categorization of all 23 sources as BL Lacs and determination of redshift estimates or lower limits for 8 sources \citep{desai2019}. The fourth paper from this campaign resulted in classification of 15 sources (out of 38) as BL Lacs using machine-learning algorithms \citep{Silver2020}.\\
There are other works that are complementary to our campaign towards completing the 3FHL catalog. For instance, \cite{paiano17,Paiano_2019}, and \cite{paiano2020} have studied 103 {\it Fermi}-LAT sources so far and provided redshifts for 65 of them. As part of a large spectroscopic program, \cite{pena2020} have classified 350 {\it Fermi}-LAT sources into BL Lac and FSRQ categories. \cite{landoni2020z} have compiled a database of spectroscopic redshifts of BL Lacs that is available online\footnote{\url{https://web.oapd.inaf.it/zbllac/}}.

We present here the results of the fifth part of this ongoing optical spectroscopic campaign aimed at the spectral completeness of the 3FHL catalog. We report on 28 sources observed with the COSMOS spectrograph mounted on the 4\,m Blanco telescope, located at the CTIO facility in Chile. Based on the optical spectra of our sources, 25 sources were classified as BL Lacs, and the remaining three were classified as FSRQs\footnote{Two of these being tentative classifications (for 3FHL J0001.9$-$4155 and 3FHL J0102.8$-$2001)}. With $\sim$85\% of the sources being BL Lacs in the 3FHL catalog, these classification results are not unexpected. \cite{desai2019} found all (23) sources observed by them to be BL Lacs, while \cite{marchesi18} identified 27 sources as BL Lacs, out of the 28 observed.

We were also able to place redshift constraints on 57\% (16/28) of our sources representing a significant improvement in the success rate of such campaigns with 4\,m class telescopes (see \citealp{landoni15}, \citealp{ricci15}, \citealp{alvarez16a}, \citealp{pena17}).

The completeness plots for our campaign sources with respect to all blazars in the energy range 10 GeV -- 2 TeV are shown in Figure~\ref{fig:compl}. The latest release of the fourth catalog of AGN detected by the \textit{Fermi}-LAT (4LAC--DR2; \citealp{Ajello2020zz}) was used to update the 3FHL catalog in order to include recent redshift measurements and classifications of sources as provided by various campaigns (e.g., \citealp{pena17,Paiano_2019,menezes2020z,paiano2020,pena2020}). Completeness was obtained by considering the ratio of the number of classified blazars or blazars with $z$ above a given flux and above divided by the total number of blazars in the same range. 
As seen from this figure, this campaign has substantially increased the type completeness of the sources in 3FHL catalog, but less so the redshift completeness. Prior to our campaign, 21\% of the 3FHL blazars lacked type and 45\% lacked a redshift value. With our campaign, we were able to decrease the fraction of blazars without type to $\sim$12\% and blazars without redshift to 43\%. The type completeness also reaches 95\% for $F>3\times10^{-12}$ \cgs (see Figure~\ref{fig:compl}, left).

Hence, even with 4\,m class telescopes like KPNO and CTIO, we have made significant progress in achieving our goal of full completeness in the 3FHL catalog. But since the success rate of redshift determination with 8\,m class telescopes is even higher (60\%$-$80\%), for the next part of our campaign, we will present the results of the observations undertaken using the 8\,m Gemini-N and -S telescopes.

\begin{figure}
    \centering
    \includegraphics[width=1.02\columnwidth]{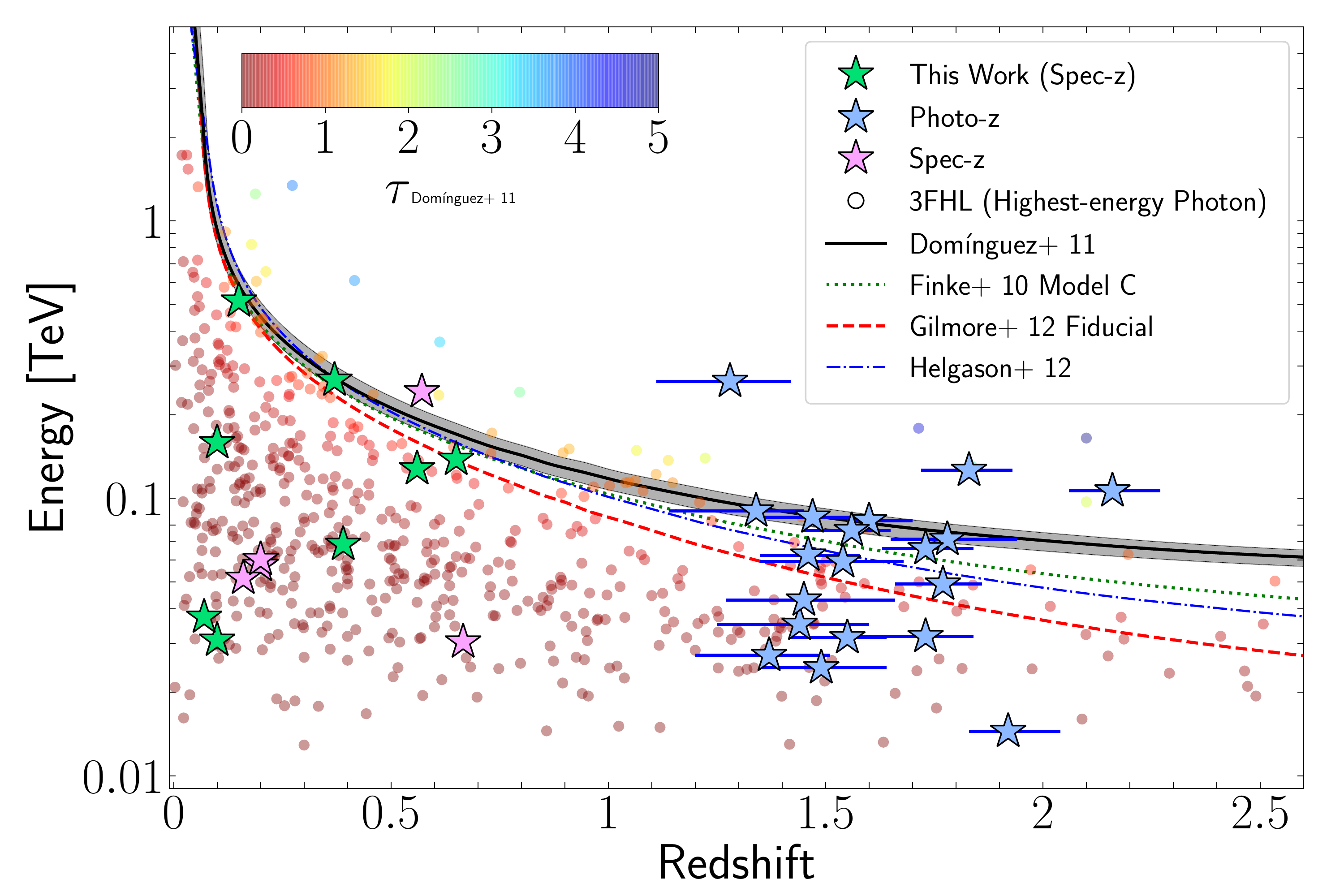}
\caption{Estimation of the Cosmic Gamma-ray Horizon as a function of redshift. The data points and some of the models have been adapted from Figure 17 of \cite{ajello17}. The highest-energy photons (HEP) from 3FHL catalog are plotted in colored circles. The color of the circles represents the corresponding optical depth values as shown in the colorbar. The prediction from different EBL models: \citealp{Finke2010} (dotted green line), \citealp{Dominguez2011} (solid black line, with uncertainties shown as the shaded band), \citealp{Gilmore2012} (dashed red line), and \citealp{Helgason2012} (dotted– dashed blue line) are shown for comparison. The pink and green stars report the HEP for sources from the spectroscopic campaign, while the blue stars represent the HEP for sources from the photometric campaign \citep{Rau2012,kaur2017,Kaur2018,Rajagopal2020}.}
    \label{fig:cgrh}
\end{figure}

Identification of classes and redshifts of high energy sources is also a crucial step in understanding and constraining the EBL (e.g., \citealp{saldana20}). The Cosmic Gamma-ray Horizon (CGRH) provides an estimate of the distance beyond which the Universe becomes opaque to $\gamma$ rays ($\tau_{\gamma\gamma}$=1, where $\tau_{\gamma\gamma}$ is the optical depth of EBL as estimated from the models). This opacity originates when very high-energy (VHE, $E>100$~GeV) photons interact with the low energy EBL photons causing electron-positron pair production to occur, thus annihilating the photons in the process \citep{dominguez13}. The CGRH as a function of redshift has been plotted in Figure~\ref{fig:cgrh} for all the highest-energy photons (HEP) in the 3FHL catalog (\citealp{ajello17}). 
We highlight in the plot both the objects studied in our spectroscopic campaign and the photometric campaign. The photometric dropout technique is based on the flux attenuation caused by the interaction of source photons by neutral Hydrogen and becomes particularly effective at z$>$1 \citep{Rau2012,kaur2017,Kaur2018,Rajagopal2020}. It is clear that both campaigns resulted in several sources for which the HEPs are very near current estimates of the CGRH, as well as some beyond the horizon\footnote{The source from the spectroscopic campaign that lies beyond $\tau>1$ is 1RXS J013427.2$+$263846, which has an HEP of $242$ GeV and $z=0.57$ \citep{marchesi18}.}. Thus, such campaigns are invaluable in further constraining the CGRH as well as probing higher opacity regions, e.g. $\tau_{\gamma\gamma}\geq$2, where redshift data are sparse. Estimating a redshift for such high-energy sources will enable better and a more accurate measurement of the EBL.
 
\section{Acknowledgements}
M.R., S.M. and M.A. acknowledge support from NASA contract 80NSSC19K0151. A.D. is thankful for the support of the Ram\'on y Cajal program from the Spanish MINECO. The authors thank Alberto Alvarez, Claudio Aguilera and Sean Points for the help provided during the observing nights at CTIO. This work made use of the TOPCAT software (Taylor 2005) for the analysis of data tables.

\begingroup
\renewcommand*{\arraystretch}{1.8}
\begin{table*}
\centering
\scalebox{0.85}{
\hspace{-2.4cm}
\begin{tabular}{lcccccccc
}
\hline
\hline
3FHL Source Name & Counterpart & R.A. J2000 & Dec J2000 & E(B-V) & AB Mag & Obs Date & Exposure & Continuum Slope\\
 & & (hh:mm:ss) & (hh:mm:ss) & (mag) & & & (sec) & \\
(1) & (2) & (3) & (4) & (5) & (6) & (7) & (8) & (9)\\
\hline
J0001.9$-$4155 & 1RXS J000135.5$-$415519 & 00:01:33.05 & $-$41:55:24.31 & 0.0114  & 18.43 & 2019 Jun 15 & 3700 & $-$2.25\\
J0102.8$-$2001 &  PMN J0102$-$2001 & 01:02:50.96 & $-$20:01:58.30 & 0.0128 & 18.20 & 2019 Jun 15 & 4350 & $-$1.91\\
J0659.5$-$6743 & SUMSS J065932$-$674346 & 06:59:32.20 & $-$67:43:46.00 & 0.1053 & 20.21 & 2019 Apr 20 & 6000 & $-$1.98\\
J0709.1$-$1525 & PKS 0706$-$15 & 07:09:12.28 & $-$15:27:00.30 & 0.5537 & 17.12 & 2019 Apr 21 & 3000 & $-$3.12\\
J0737.5$-$8247 & SUMSS J073706$-$824836 & 07:37:06.80 & $-$82:48:37.00 & 0.1606 & 17.63 & 2019 Apr 21 & 3600 & $-$1.27\\
J0953.3$-$7659 & SUMSS J095303$-$765804 & 09:53:04.38 & $-$76:58:02.20 & 0.4551 & 18.69 & 2019 Apr 19 & 5100 & $-$1.85\\
J1042.2$-$4128 & 1RXS J104204.1$-$412936 & 10:42:04.09 & $-$41:29:35.99 & 0.0774 & 18.36 & 2019 Jun 13 & 4000 & $-$4.92\\
J1130.7$-$3137 & NVSS J113046$-$313805 & 11:30:46.10 & $-$31:38:08.00 & 0.0533 & 16.07 & 2019 Apr 20 & 1800 & $+$1.59\\
J1225.4$-$3447 & 1RXS J122534.0$-$344737 & 12:25:34.00 & $-$34:47:37.00 & 0.0549 & 18.36 & 2019 Jun 12 &  4100 & $-$3.94\\
J1328.9$-$5607 & PMN J1329$-$5608 & 13:29:01.14 & $-$56:08:02.65 & 0.4101 & 16.78 & 2019 Apr 19 & 1800 & $-$2.78\\
J1339.0$-$2359 & PKS 1336$-$237 & 13:39:01.70 & $-$24:01:14.00 & 0.0678 & 17.49 & 2019 Apr 20 & 3000 & $-$2.89\\
J1353.8$-$3936 & NVSS J135345$-$393711 & 13:53:45.16 & $-$39:37:11.40 & 0.0571 & 19.30 & 2019 Apr 21 & 5400 & $-$0.18\\
J1442.5$-$4621 & SUMSS J144236$-$462302 & 14:42:36.40 & $-$46:23:02.00 & 0.1271 & 16.40 & 2019 Apr 19 & 3600 & $-$2.46\\
J1455.4$-$7559 & SUMSS J145543$-$760054 & 14:55:43.23 & $-$76:00:54.61 & 0.1213 & 18.40 & 2019 Jun 12 & 5200 & $-$1.10\\
J1457.8$-$4642 & PMN J1457$-$4642 & 14:57:41.80 & $-$46:42:10.00 & 0.1990 & 16.63 & 2019 Apr 20 & 1800 & $+$0.41\\
J1514.7$-$0949 & PMN J1514$-$0948 & 15:14:49.60 & $-$09:48:38.00 & 0.0917 & 19.48 & 2019 Jun 13 & 4800 & $-$1.28\\
J1542.1$-$2915 & NVSS J154203$-$291509 & 15:42:03.10 & $-$29:15:09.00 & 0.1553 & 18.72 & 2019 Jun 12 & 4900 & $-$2.27\\
J1549.7$-$3045 & NVSS J154946$-$304501 & 15:49:46.60 & $-$30:45:01.00 & 0.1026 & 19.25 & 2019 Jun 13 & 5400 & $-$1.60\\
J1604.6$-$4441 & PMN J1604$-$4441 & 16:04:31.02 & $-$44:41:31.97 & 0.7779 & 20.00 & 2019 Jun 15 & 6200 & $-$0.92\\
J1647.3$-$6438 & PMN J1647$-$6437 & 16:47:37.74 & $-$64:38:00.26 & 0.1522 & 18.15 & 2019 Apr 20 & 3700 & $-$0.62\\
J1849.2$-$1647 & 1RXS J184919.7$-$164726 & 18:49:19.70 & $-$16:47:26.00 & 0.2714 & 18.43 & 2019 Jun 13 & 2600 & $-$1.84\\
J1934.2$-$2419 & NVSS J193412$-$241922 & 19:34:12.69 & $-$24:19:19.92 & 0.0939 & 17.26 &  2019 Jun 15 & 3400 & $-$1.58\\
J1944.4$-$4523 & 1RXS J194422.6$-$452326 & 19:44:22.39 & $-$45:23:30.12 & 0.0450 & 15.62 & 2019 Jun 13 & 900 & $-$1.57\\
J1944.9$-$2143 & 1RXS J194455.3$-$214318 & 19:44:55.10 & $-$21:43:18.84 & 0.0733 & 18.06 & 2019 Jun 12 & 3900 & $-$2.48\\
J1955.0$-$5640 & 1RXS J195503.1$-$564031 & 19:55:03.00 & $-$56:40:31.08 & 0.0453 & 17.25 & 2019 Jun 12 & 3100 & $-$1.98\\
J2002.5$-$7119 & 1RXS J200234.9$-$711943 & 20:02:27.12 & $-$71:19:40.80 & 0.0607 & 19.68 & 2019 Jun 12 & 3300 & $-$1.81\\
J2058.8$-$1442 & TXS 2056$-$149 & 20:58:46.74 & $-$14:43:05.00 & 0.0351 & 15.38 & 2019 Jun 13 & 900 & $-$0.75\\
J2309.7$-$3633 & 1RXS J230940.6$-$363241  & 23:09:40.85 & $-$36 32 49.00 & 0.2731 & 17.75 & 2019 Jun 12 & 3150 & $-$3.44\\
\hline 

\hline
\hline
\end{tabular}}\caption{\label{tab:tabulated1} List of sources and their properties sorted in the order of increasing R.A. (Right ascension) values. The columns are: (1) 3FHL catalog \citep{ajello17} name for the source. (2) optical, IR, X-ray or radio counterpart of the source, (3) Right ascension, (4) Declination, (5) $E(B-V)$ value obtained using the measurements of \citet{schlafly11} and the NASA/IPAC Infrared Science Archive online tool, (6) V band magnitude (AB system), (7) Date of observation, (8) Exposure time (in seconds) of the source under consideration, and (9) Slope of continuum fit obtained from the observed spectra.}
\label{tab:sample}
\end{table*}
\endgroup



\clearpage
\startlongtable
\begin{deluxetable*}{cccccccc}
\tablecolumns{6}
	\tablecaption{\label{tab:redshift} Results obtained from spectral analysis discussed in Section~\ref{results}. Classifications and redshift measurements marked with * are tentative results.}
	\tablewidth{2pt}
	\tabletypesize{\small}
	\setlength{\tabcolsep}{0.06in} 
	\tablehead{
	    \colhead{3FHL Source Name} & \colhead{S/N} & \colhead{Spectral line} & \colhead{Observed $\lambda$ (\r{A})} & \colhead{Line type} & \colhead{Equivalent Width (\AA)} & \colhead{Redshift} & \colhead{Classification}\\
 & & \colhead{Rest frame $\lambda$ (\r{A})} & & & &}
 
\startdata
J0001.9$-$4155 & 25.36 & -- & 5450 & Emission & 88 & -- & FSRQ*\\
J0102.8$-$2001 & 18.40 & O {\sc ii} (3727) & 5107 & Emission  & 18 & 0.37& FSRQ*\\
& & NaD (5892) & 8084 & Absorption & & &\\
J0659.5$-$6743 & 15.97 & & & & & & BL Lac\\
J0709.1$-$1525 & 39.40 & H$\beta$ (4861) & 5551 & Absorption & & $>$0.14 & BL Lac\\
& & Ca$+$Fe (5269) & 6017 & Absorption & & &\\
J0737.5$-$8247 & 30.37 & & & & & & BL Lac\\
J0953.3$-$7659 & 13.94 & & & & & & BL Lac\\
J1042.2$-$4128 & 24.07 & G$-$band (4304) & 5210 & Absorption & & $>$0.21 & BL Lac\\
& & Mg {\sc i} (5175) & 6274 & Absorption & & &\\
J1130.7$-$3137 & 39.41 & H$\beta$ (4861) & 5609 & Emission & 2.4 & 0.15 & BL Lac\\
& & O {\sc iii} (5007) & 5770 & Emission & 1.1 & & \\
J1225.4$-$3447 & 29.84 & G$-$band (4304) & 5994 & Absorption & & $>$0.39 & BL Lac\\
& & Ca$+$Fe (5269) & 7336 & Absorption & & &\\
J1328.9$-$5607 & 20.21 & G$-$band (4304) & 5063 & Absorption & & $>$0.17 & BL Lac\\
& & Mg {\sc i} (5175) & 6099 & Absorption &  & &\\
J1339.0$-$2359 & 23.29 & O {\sc ii} (3727) & 6165 & Emission & 47 & 0.65 & FSRQ\\
& & O {\sc iii} (5007) & 8287 & Emission & 22 & &\\
J1353.8$-$3936 & 29.88 & G$-$band (4304) & 6440 & Absorption & & $>$0.49 & BL Lac\\
& & Ca$+$Fe (5269) & 7875 & Absorption & & &\\
J1442.5$-$4621 & 64.63 & O {\sc iii} (5007) & 5517 & Emission & 1.5 & 0.10 & BL Lac\\
& & Mg {\sc i} (5175) & 5698 & Absorption & & &\\
& & Ca$+$Fe (5269) & 5806 & Absorption & & &\\
& & NaD (5892) & 6492 & Absorption & & &\\
J1455.4$-$7559 & 13.52 & & & & & & BL Lac\\
J1457.8$-$4642 & 36.07 & H$\beta$ (4861) & 5372 & Emission & 0.6 & 0.10 & BL Lac\\
& & Mg {\sc i} (5175) & 5744 & Absorption & & &\\
& & Ca$+$Fe (5269) & 5851 & Absorption & & &\\
J1514.7$-$0949 & 21.56 & & & & & & BL Lac\\
J1542.1$-$2915 & 20.22 & & & & & & BL Lac\\
J1549.7$-$3045 & 12.38 & & & & & & BL Lac\\
J1604.6$-$4441 & 06.39 & & & & & & BL Lac\\
J1647.3$-$6438 & 43.31 & Ca {\sc ii} (3934) & 7195 & Absorption & & $>$0.82 & BL Lac\\
& & Ca {\sc ii} (3969) & 7230 & Absorption & & &\\ 
J1849.2$-$1647 & 68.07 & Mg {\sc i} (5175) & 8537 & Absorption & & $>$0.64 & BL Lac\\
& & Ca$+$Fe (5269) & 8659 & Absorption & & & \\
& & OR & & & & &\\
& & Mg {\sc i} (5175) & 8659 & Absorption & & $>$0.67 & \\
& & Ca$+$Fe (5269) & 8805 & Absorption & & & \\
J1934.2$-$2419 & 26.89 & & & & & & BL Lac\\
J1944.4$-$4523 & 34.53 & G$-$band (4304) & 7274 & Absorption & & $>$0.68 & BL Lac\\
& & Mg {\sc i} (5175) & 8704 & Absorption & & &\\
& & Ca$+$Fe (5269) & 8916 & Absorption & & &\\
J1944.9$-$2143 & 20.29 & Mg {\sc i} (5175) & 7289 & Absorption & & $>$0.41 & BL Lac\\
& & Ca$+$Fe (5269) & 7435 & Absorption & &\\
J1955.0$-$5640 & 25.47 & G$-$band (4304) & 5255 & Absorption & & $>$0.22 & BL Lac\\
& & NaD (5892) & 7198 & Absorption & & &\\
J2002.5$-$7119 & 23.10 & & & & & & BL Lac\\
J2058.8$-$1442 & 35.05 &  O {\sc iii} (5007) & 5375 & Emission & 0.4 & 0.07 & BL Lac\\
& & Mg {\sc i} (5175) & 5574 & Absorption & & & \\
J2309.7$-$3633 & 43.26 & & & & & & BL Lac\\
\enddata
\end{deluxetable*}

\clearpage

\bibliographystyle{aasjournal}
\bibliography{3FHL_ctio}

\end{document}